\documentstyle[psfig,epsf,nat_julie,pslatex,afterpage,tabularx]{mn2e}
\begin{document}

\newif\ifAMStwofonts


\title[The depolarisation properties of powerful radio sources.]
{The depolarisation properties of powerful radio sources: breaking the radio-power vs. redshift degeneracy.}

\author[J.A.Goodlet et al.]
      {J.A.Goodlet$^{1}$\thanks{email: jag@astro.soton.ac.uk}, C. R. Kaiser$^{1}$, P.N.Best$^2$, J. Dennett-Thorpe$^3$\\
$^1$ Department of Physics \& Astronomy, University of Southampton, Southampton SO17 1BJ\\
$^2$ Institute for Astronomy, Royal Observatory Edinburgh, Blackford Hill, Edinburgh EH9 3HJ\\
$^3$ Astron, P.O. Box 2, 7990 AA Dwingeloo, the Netherlands}

\date{Accepted ??????.
      Received ??????;
      in original form 2002 May}

\pagerange{\pageref{firstpage}--\pageref{lastpage}}
\pubyear{2002}

\maketitle

\label{firstpage}

\begin{abstract}
We define 3 samples of extragalactic radio sources of type FRII,
containing 26 objects in total. The control sample consists of 6C and
7C sources with radio powers of around $10^{27}$\,W\,Hz$^{-1}$ at
151\,MHz and redshifts of $z\sim 1$. The other samples contain 3CRR
sources with either comparable redshifts but radio powers about a
decade larger or with comparable radio powers but redshifts around
$z\sim 0.4$. We use these samples to investigate the possible
evolution of their depolarisation and rotation measure properties with
redshift and radio power independently. We used VLA data for all
sources at $\sim$ 4800 MHz and two frequencies within the 1400
MHz band, either from our own observations or from the archive. We present
maps of the total intensity flux, polarised flux, depolarisation,
spectral index, rotation measure and magnetic field direction where
not previously published. Radio cores were detected in twelve of the
twenty-six radio sources. Fourteen of the sources show a strong
Laing-Garrington effect but almost all of the sources show some
depolarisation asymmetry. All sources show evidence for an external
Faraday screen being responsible for the observed depolarisation.
We find that sources at higher redshift are more strongly
depolarised. Rotation measure shows no trend with either redshift or
radio power, however variations in the rotation measure across
individual sources increase with the redshift of the sources but do
not depend on their radio power.
\end{abstract}
\section{Introduction}
\renewcommand\dbltopfraction{.9}
\renewcommand\bottomfraction{.5}
\renewcommand\textfraction{.2}
\setcounter{totalnumber}{500}
\setcounter{topnumber}{500}
\setcounter{bottomnumber}{500}
Observations of the polarisation properties of extragalactic
radio sources can provide information on the relationships between the
radio source properties and their environments as well as the
evolution of both with redshift. Many previous studies of variations in
polarisation properties have suffered
from a degeneracy between radio power and redshift due to Malmquist
bias, present in all flux-limited samples. A good example of this
effect is the depolarisation correlations found independently by
\citet{kcg72} and \citet{mt73}.

\citet{kcg72} found that depolarisation of the radio lobes generally
increased with redshift whereas \citet{mt73} found depolarisation to
increase with radio luminosity. Due to the flux-limited samples (PKS and 3C)
used by both authors it is difficult to distinguish which is the
fundamental correlation, or whether some combination of the two
occurs. Both suggestions have ready explanations: (i) If radio
sources are confined by a dense medium then synchrotron losses due to
adiabatic expansion are reduced, the internal magnetic field is
stronger and a more luminous radio source results; if this confining
medium also acts as a Faraday medium, more luminous sources will tend
to be more depolarised. (ii) Sources at different cosmological epochs
may reside in different environments and/or their intrinsic properties
may change with redshift.

\citet{hl91} observed that galaxy densities around FR II radio sources
increased with redshift out to $z \approx 0.5$ and beyond but
\citet{wls01} did not find this trend in a recent study.
\citet{wpk84} argued that the increase in rotation measure with
redshift is primarily attributable to an increasing contribution of
intervening matter. However, depolarisation asymmetries within a
source, e.g the Laing-Garrington effect, increase with redshift which
imply an origin local to the host galaxy \citep{gc91a}.

\begin{figure}
\centerline{\psfig{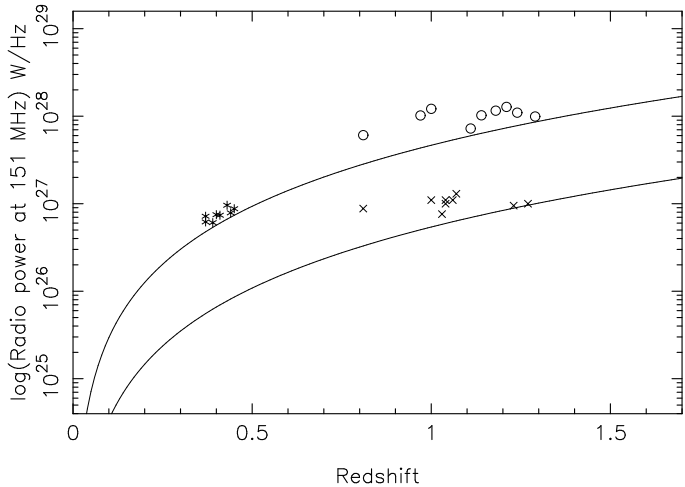}}
\caption{A radio power-redshift plot showing the 3 subsamples used in
the observations. Sample A is represented by `x', sample B by `o' and
sample C by `*'. The lines mark the flux limits for the 3CRR and 6C
samples. A spectral index of 0.75 was used to shift the 3CRR data to
151 MHz.}\label{pzplot}
\end{figure}

To break this apparent degeneracy effect we defined 3 subsamples of sources chosen from
the 3CRR and 6C/7C catalogues: The control sample consists of 6C and 7C sources
at redshift $z \approx 1$ with radio powers of around $10^{27}$
WHz$^{-1}$ at 151 MHz. Another sample at the same redshift
consists of 3CRR sources with radio powers around a magnitude higher
at 151 MHz. The final sample consists of 3CRR sources at redshift $z
\approx 0.4$, again  with radio powers of around $10^{27}$ WHz$^{-1}$ at
151 MHz (Figure \ref{pzplot}).
The observations of sample sources can then be
used to study the source properties and the medium around the
source, thus discovering which correlate with redshift and which
correlate with radio power enabling us to answer the following
questions:
\begin{center}
\begin{itemize}
\item Does a relationship exist between radio power and the \\
	\hspace{0.2cm} environment in which a given radio source lives?

\item Do the environments evolve with redshift?

\end{itemize}
\end{center}
\begin{table}
\centering \caption{Details of the sources in sample A,B \& C. Sources in italics are quasars. $\sigma _{\nu}$ gives the noise level in the final total flux maps at frequency $\nu$.} \label{setup}
\begin{tabular}{cccccc}
Source & z&P$_{151 MHz}$   & Angular size & $\sigma_{4.8}$ & $\sigma_{1.4}$\\
& &(W/Hz)  &  (arcsec)   & ($\mu$ Jy) & ($\mu$ Jy)\\
6C0943+39&1.04&$1.0\times10^{27}$&10&20&55\\
6C1011+36&1.04&$1.1\times10^{27}$&49&15&65\\
6C1018+37&0.81&$8.8\times10^{26}$&64&24&50\\
6C1129+37&1.06&$1.1\times10^{27}$&15&21&60\\
6C1256+36&1.07&$1.3\times10^{27}$&14&18&70\\
6C1257+36&1.00&$1.1\times10^{27}$&38&33&100\\
{\it 7C1745+642}&1.23&$9.5\times10^{26}$&16&30&60\\
{\it 7C1801+690}&1.27&$1.0\times10^{27}$&21&24&75\\ \vspace{0.3cm}
{\it 7C1813+684}&1.03&$7.1\times10^{26}$&52&23&47\\

3C65&1.18&$1.0\times10^{28}$&17&22&61\\
{\it 3C68.1}&1.24&$1.1\times10^{28}$&52&41&100\\
3C252&1.11&$7.2\times10^{27}$&60&23&100\\
3C265&0.81&$6.5\times10^{27}$&78&32&70\\
3C268.1&0.97&$1.0\times10^{28}$&46&32&68\\
3C267&1.14&$1.0\times10^{28}$&38&25&61\\
3C280&1.00&$1.2\times10^{28}$&15&33&75\\
3C324&1.21&$1.3\times10^{28}$&10&26&70\\\vspace{0.3cm}
{\it 4C16.49}&1.29&$9.9\times10^{27}$&16&25&81\\

3C16&0.41&$7.4\times10^{26}$&63 & 36& 80\\
3C42&0.40&$7.5\times10^{26}$&28&29 &70\\
3C46&0.44&$7.9\times10^{27}$&168&20&53\\
3C299&0.37&$6.9\times10^{26}$&11&39&77\\
3C341&0.45&$8.8\times10^{26}$&70&26&88\\
{\it 3C351}&0.37&$7.6\times10^{26}$&65&31&100\\
3C457&0.43&$9.6\times10^{27}$&190&18&80\\\vspace{0.3cm}
4C14.27&0.39&$8.8\times10^{26}$&30&30&90\\
\end{tabular}
\end{table}

The structure of the paper is as follows. Section 2 describes in
detail the sample selection and the VLA observations. Section 3
contains the results, including the maps of the 26 sources from
the 3 samples. Section 4 discusses the observed trends across the
samples, detailing any correlations between observables. Section 5
summarises the conclusions of the paper. All values are calculated assuming $H_o= 50$
kms$^{-1}$Mpc$^{-1}$, and $\Omega_m = 0.5$ ($\Lambda = 0$).

\section{The VLA observations and data reduction}
\subsection{Sample selection}
\begin{figure*}
\centerline{\psfig{file=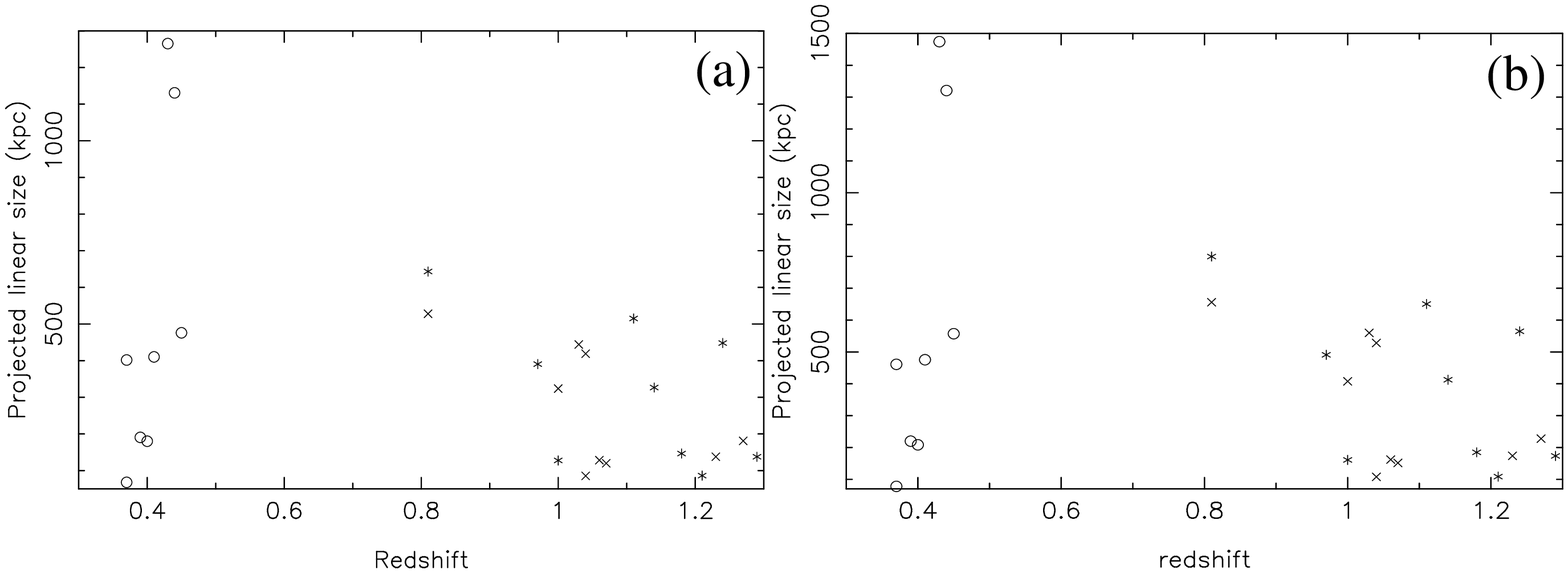,width=18cm}}
\caption{Linear size-redshift plots of the 3 subsamples used in the observations. Symbols as in Figure \ref{pzplot}. Figure (a) assumes $H_o= 50$
kms$^{-1}$Mpc$^{-1}$, and $\Omega_{m} = 0.5$, $\Omega_{\Lambda} = 0$. Figure (b) assumes $H_o= 50$
kms$^{-1}$Mpc$^{-1}$, and $\Omega_{m} = 0.35$, $\Omega_{\Lambda} = 0.65$.}\label{sizeplot}
\end{figure*}

Sample A was defined as a subsample chosen from the 6CE \citep{erl97}
subregion of the 6C survey \citep{hmw90}, and the 7C III subsample
\citep{lrh99}, drawn from the 7C and 8C surveys \citep{pwr98}. The
selected sources have redshifts $0.8<z<1.3$, and radio powers at 151
MHz between $6.5 \times 10^{26} {\rm WHz^{-1}}<P_{151 {\rm MHz}}<1.35
\times 10^{27} {\rm WHz^{-1}}$. Sample B was defined as a subsample
from the revised 3CRR survey by \citet{lrl83} containing sources within the same
redshift range but with powers in the range $6.5 \times 10^{27} {\rm
WHz^{-1}}<P_{151 {\rm MHz}}<1.35 \times 10^{28} {\rm
WHz^{-1}}$. Sample C is also from the 3CRR catalogue; it has the same
radio power distribution as the control sample, sample A, but with
$0.3<z<0.5$.  We only include sources that were more luminous
than the flux limits of the original samples at 151 MHz. (Figure
\ref{pzplot}). In all samples only sources with angular sizes $
\theta\ge 10''$ (corresponding to $\approx$ 90 kpc at z=1) were
included, see Table \ref{setup}. This angular size
limit is imposed by the depolarisation measurements as they require a
minimum of ten independent telescope beams (1'' per beam) over the
entire source. The distributions of linear sizes of the radio lobes
are reasonably matched across all the samples (Figure
\ref{sizeplot}). Each sample initially contained 9 sources; the source
3C109 was subsequently excluded from sample C as the VLA data was of
much poorer quality than that of the rest of the sample. Each of the
resulting subsamples contains 1 to 3 quasars.  The sources in the 3
subsamples are representative of sources with similar redshifts, radio
powers and sizes. However, the samples are not statistically complete
because of observing time limitations. We picked sources that were
fairly well observed at 4.8 GHz at B array (and C array if needed) in
the archives, ensuring that a minimum of new observations was
needed. Full details of the samples are given in Table \ref{setup}

The ratio of angular to physical size varies only by a factor 1.3
between $z=0.4$ and $z=1.4$. This ensured that all the sources
were observed at similar physical resolutions. 
All redshift values for 6C sources were taken from
\citet{erl97} except for 6C1018+37 which was taken from \citet{rel01}. The 7C sources 
were taken from \citet{lrh99}.
The 3CRR sources were taken from \citet{sdm97},
4C16.49 was taken from \citet{strange},
4C14.27 was taken from \citet{hr92} and 3C457 was taken from \citet{hb91}.
\subsection{Very Large Array observations}
\label{obs}
Observations of all 26 radio galaxies were made close to 1.4 GHz
using the A array configuration and a 25 MHz bandwidth. This
bandwidth was used instead of 50 MHz to reduce the effect of
bandwidth depolarisation. The maximum angular size that can be
successfully imaged using A array at 1.4 GHz is 38$''$. 12 sources
were larger than this and were observed additionally with B array.
Observations were also made at 4.8 GHz using a 50 MHz bandwidth.
The maximum observable angular size in B array at 4.8 GHz is 36$''$.
The same 12 sources as before, were then observed at 4.8 GHz with
C array. This ensured that both 1.4 GHz and 5 GHz observations
were equally matched in sensitivity and resolution.
Details of the observations are given in Tables \ref{vlaobs} to
\ref{vlaobsc}.

Sources in the 6C and 7C samples have a typical bridge surface
brightness of $\approx 70 \mu$Jy beam$^{-1}$ in 5 GHz A-array
observations \citep{bel99}. In order to detect 10\% polarisation at
$3\sigma$ in B-array observations we required an rms noise level of
$20 \mu$Jy beam$^{-1}$, corresponding to 70 mins of integration time.
At 1.4 GHz, assuming $\alpha =-1.3$, bridges will be a factor of 4
more luminous. At this frequency,
the exposure time is set by the requirement to have an adequate amount
of {\it uv\/}-coverage to map the bridge structures. 20 minute observations
were split into 4$\times$5 minute intervals. This observation splitting
to improve {\it uv\/}-coverage was also done for the 4.8 GHz data. 

At 1.4 GHz the
integration time on all the sources is above the minimum required for
good signal to noise.As Table \ref{vlaobs} demonstrates, for many of our sources the
integration times at 5 GHz are considerably less than the 70 min
requirement, due to telescope time constraints Many of the
observed properties that depend on polarisation observations (e.g
depolarisation and rotation measure), are therefore poorly
measured in the fainter components at 5 GHz.  The values
obtained are then only representative of the small region detected and
not the entire component. Spectral index is independent of the
polarisation measurements and so it is relatively unaffected by the
short integration times.

The 3CRR sources are more luminous, but much of this is due to the
increase in the luminosity of their hotspots; their bridge
structures are only a few times brighter than those of the 6C/7C
III sources. To reach a $3\sigma$ detection of 7\% polarisation on
the bridge structures, a total integration time of 30 mins was required
at 5 GHz and 20 mins at 1.4 GHz, split into 3-4 minute intervals
to improve {\it uv\/}-coverage. The vast majority of the sources
in sample B and C had at least this minimum amount of time on
source ( see Tables \ref{vlaobsb} and \ref{vlaobsc}).

Most observations at 1.4 GHz using A array were obtained on
31/07/99 (AD429). The data from this day is strongly affected by a
thunderstorm at the telescope site during most of the
observations. Even after removal of bad baselines and antennas the
noise level in this data remained at least twice that of the
theoretical value. However, careful calibration and {\small CLEAN}ing
reduced this effect to a minimum. Sample A was most affected by the
thunderstorm and the lack of observing time at all
 frequencies. However, we find that the results obtained by
 \citet{bel99} for some of the sources in sample A are in good
 agreement with our results. We are therefore confident that our
 data is reliable for fluxes above the 3$\sigma _{\rm noise}$ level.
We checked the polarisation calibration of 31/07/99 (AD429) by comparing with 
the B-array data at 1.4 GHz (for the 12 sources that had B-array data) to confirm
that the PA of both data sets agreed to within 15 degrees in all sources.
This additional check allowed us to ensure that the 1.4 GHz polarisation angle
calibration was accurate. 

The data were reduced using the {\small AIPS} software package
produced by the National Radio Astronomy Observatory. At 1.4 GHz the
two IFs were reduced separately producing independent observations at
1665 MHz and 1465 MHz.  The reason for the separation is due to the
significant rotation of the polarisation angle between the two
frequencies. If the two frequencies were not separated then there
would be some degree of artificial depolarisation at 1.4 GHz; this was
not a problem at 4.8 GHz. Each source was then {\small CLEAN}ed using
{\small IMAGR}, an {\small AIPS} task, and then improved by two cycles
of phase self-calibration followed by amplitude-phase
self-calibration. For sources larger than 12$''$ the {\it uv\/}-data
from the low resolution array configurations were self-calibrated
using source models resulting from the high resolution arrays to
eliminate positional discrepancies. The combined dataset was then
cycled through another round of amplitude-phase calibration.

In many cases we used VLA archive data. Table
\ref{maps} lists articles containing this previously published
data. To maintain the consistency of our samples we re-analysed
all the data.

\begin{table}
\centering \caption{Details of the VLA observations for sample A
with the integration times included. See Tables \ref{vlaobsb} and
\ref{vlaobsc} for samples B and C  respectively.} \label{vlaobs}
\begin{tabular}{p{0.85cm}ccp{0.6cm}cp{0.5cm}}
Source & Array  & Frequency & Bandwidth& Observing & Int. \\
& Config. & (MHz)&(MHz)& Program  & (min)\\
6C0943+39 & A &1465,1665&25  & 31/07/99  (AD429) & 16 \\
&  B & 4885,4535&50& 20/05/01 (AD444)&31\\
6C1011+36 & A &1465,1665&25 & 31/07/99 (AD429) & 16\\
&  B & 1452,1652&25&  20/05/01 (AD444)& 17\\
&  B & 4885,4535&50 & 25/02/97 (AL397)&21\\
&  &4885,4535&50 & 20/05/01 (AD444)&30 \\
&  C & 4885,4535&50 & 12/06/00 (AD429)& 20 \\
6C1018+37  & A & 1465,1665 & 25&31/07/99 (AD429)& 16\\
&  B & 1452,1652&25 & 20/05/01 (AD444)&17\\
&  B & 4885,4535&50 & 20/05/01 (AD444)&31\\
&  C & 4885,4535&50 & 12/06/00(AD429)& 20 \\
6C1129+37  & A & 1465,1665&25  & 31/07/99 (AD429)& 16\\
&  B & 4885,4535&50  & 20/05/01 (AD444)& 17\\
&  & 4885,4535&50 &  25/02/97 (AL397)& 21\\
6C1256+36  & A &1465,1665&25  & 31/07/99 (AD429)& 16\\
&  B & 4885,4535&50 & 27/02/93 (AR287)&15\\
6C1257+36 & A &1465,1665&25  & 31/07/99 (AD429)& 16\\
&  B & 4885,4535&50  &20/05/01 (AD444)& 16\\
&  & 4885,4535&50 & 25/02/97 (AL397)& 22\\
7C1745+642  & A &1465,1665&25  & 31/07/99 (AD429)& 16\\
&  B & 4885,4535&50  &20/05/01 (AD444)&11 \\
&  & 4885,4535&50 &  23/11/97 (AL401)& 31\\
7C1801+690  & A &1465,1665&25  & 31/07/99 (AD429)&16 \\
&  B & 4885,4535&50  &26/03/96 (AB978)& 29\\
&  & 4885,4535&50 &  23/11/97 (AL401)& 17\\
7C1813+684  & A & 1465,1665&25  & 31/07/99 (AD429)& 16\\
&  B & 1452,1652&25  & 20/05/01 (AD444)&16 \\
&  B & 4885,4535&50  & 20/05/01 (AD444)&39\\
&  & 4885,4535&50 & 23/11/97 (AL401)& 19\\
&  C &4885,4535&50  & 12/06/00 (AD429)& 20\\
\end{tabular}
\end{table}
\begin{table}
\caption{Details of the VLA observations for sample B} \label{vlaobsb}
\begin{tabular}{p{0.5cm}ccccp{0.5cm}}
Source &Array  & Frequency &Bandwidth& Observing &Int. \\
&  Config. & (MHz)&(MHz)& Program& (min) \\
3C65& A &1465,1665&25 & 31/07/99 (AD429)& 16\\
&  B &4885,4535&50  & 20/05/01 (AD444) &20\\
3C68.1  & A & 1417,1652&25  & 31/07/99 (AD429)& 16 \\
&  B &1417,1652&25  & 13/07/86 (AL113) &20\\
&  B &4885,4535&50  & 19/07/86 (AB369)& 300\\
&  C &4885,4535&50  & 12/06/00 (AD429) &20\\
3C252  & A & 1465,1665&25 & 31/07/99 (AD429)& 16 \\
&  B &1465,1665&25  & 20/05/01 (AD444) & 27\\
&  B &4885,4535&50  & 19/07/86 (AB369)&97\\
&  C & 4885,4535&50  & 12/06/00 (AD429) &20\\
3C265  & A & 1417,1652&25  & 31/07/99 (AD429)&16 \\
&  B & 1417,1652&25  & 13/07/86 (AL113) &30\\
&  B & 4873,4823&50  & 17/12/83 (AM224)& 238\\
&  C& 4873,4823&50 & 12/06/00 (AD429) &20\\
3C267  & A & 1465,1665&25  & 31/07/99 (AD429)&16 \\
&  B & 4873,4823&50 & 17/12/83 (AM224)&56 \\
3C268.1  & A & 1417,1652&25  & 31/07/99 (AD429)&16\\
 &  B & 1417,1652&25  & 13/07/86 (AL113) & 30\\
&  B & 4885,4835&50  & 15/08/88 (AR166)& 20\\
&  & 4885,4835&50 & 01/06/85 (AR123) &21\\
&  C & 4885,4835&50  & 06/11/86 (AL124)&102\\
3C280  & A & 1465,1665&25 & 31/07/99 (AD429)&16 \\
&  B & 4873,4823&50 & 17/12/83 (AM224) &46\\
3C324  & A & 1465,1665&25 & 31/07/99 (AD429)&16 \\
&  B & 4873,4823&50 & 17/12/83 (AM224)&51 \\
4C16.49  & A &1465,1652&25 & 31/07/99 (AD429)&16 \\
& B & 4885,4535&50  & 04/03/97 (AB796)&30\\
\end{tabular}
\end{table}
\subsection{Map production}
\label{blurb}
Total intensity maps were made from the Stokes I parameters at each
frequency.  Polarisation maps were also made at all frequencies by
combining the Stokes Q and U polarisation parameters. A map was then
produced that contained the polarised flux, ${\rm P}=(Q^2+U^2)^{1/2}$ and the electric field
position angle, ${\rm PA}=0.5tan^{-1}(\frac{U}{Q})$ at a given frequency.
The {\small AIPS} task
{\small POLCO} was used to correct for Ricean bias, which arises when the Stokes Q
and U maps are combined without removing noise-dominated
pixels. By careful setting of the 
{\small PCUT} parameter this bias was removed.  All maps only contain
pixels where the polarised flux and the total intensity flux are above
$5\sigma_{noise}$ at 4.8 GHz and $3\sigma_{noise}$ at 1.4 GHz. 
The lower threshold at 1.4 GHz was necessary because the 1.4 GHz data had
a higher noise level, so blanking flux below $5\sigma$ resulted in
large regions of polarised flux being lost.

At all frequencies the individual maps were made such that the beam
size, the cell size of the image and the coordinates of the
observations were exactly the same.  If any of these properties of the
map differed between frequencies then the resultant multi-frequency
map would contain false structures that would be directly related to
the mis-alignment of the maps. To make sure that the coordinates (and
cell size) were always within acceptable tolerances the {\small AIPS}
task {\small HGEOM} was used to realign maps at one frequency
to maps at another frequency. In sources where an identifiable core
exists at both frequencies, the core positions were used as a check on the
alignment from {\small HGEOM}.  In general 
{\small HGEOM} is adequate in aligning the multi-frequency data.
Sources with a distinct core at all frequencies were aligned within
0.03'', where no core existed the hotspots were aligned within 0.045''. In 4 sources this was 
not sufficient. 3C68.1 had to be shifted 0.05'' east and 0.07'' north, 3C265 had to be shifted 0.04''
west and 0.02'' north, 3C299 had to be shifted 0.1'' east and 0.03''
north and finally 3C16 had to be shifted 0.1'' east and 0.1'' north. 
All shifts were applied to the 5 GHz observations.

We define the spectral index , $\alpha$, by $S_{\nu} \propto \nu^{\alpha}$.
Spectral index maps were made between 4.8 GHz and 1.4 GHz.

Depolarisation depends on the amount of polarised flux at two
frequencies but also on the total intensity flux at the same two
frequencies. It is defined as:
\begin{equation}\label{dp}
DM^{4.8}_{1.4} = \frac{ PF_{4.8}}{PF_{1.4}},
\end{equation}
where the $PF_{\nu}$ is the fractional polarisation, (polarised
flux)/(total intensity flux) at a given frequency, $\nu$. We derived
depolarisation maps for each source. The depolarisation values given
in Tables \ref{A} to \ref{C} are average values of the
depolarisation of source components.

Rotation measure is related to the degree of rotation of the
polarisation position angle over a set frequency range, in our case
from 1.4 GHz to 4.8 GHz. The rotation measure, RM, in the
observers frame of reference, depends on the electron density, $n_e$, the component of the magnetic field along the line of sight,
$B_{\|}$, the wavelength of the observations, $\lambda$, and
the path length, l.
\begin{equation}
RM = 8.1\times10^{3}\int_0^D{n_eB_{\|} dl} \hspace{0.5cm} {\rm rad}\hspace{0.1cm}{\rm m}^{-2} ,
\end{equation}
\begin{equation}
PA(\lambda) = PA_{\circ} + RM \lambda^2,
\end{equation}
where PA is the observed polarisation position angle of a source
and PA$_{\circ}$ is the initial polarisation angle before any
rotation.
Three frequencies (1.4 GHz, 1.6 GHz and 4.8 GHz) were used to overcome
the $n\pi$ ambiguities when fitting to the observed polarisation
angles \citep{Simard,Rudnick}. Thus depolarisation measurements depend on
observations at only 2 frequencies whereas rotation measure depends on
data at all 3 observed frequencies, (i.e. considering the two 1.4 GHz
IFs separately), making it more sensitive to the level of polarisation
observed in a source.  This meant that in some cases (e.g. 3C16,
Figure \ref{3c16}) there were depolarisation measurements in one lobe
but there was no corresponding rotation measure. 

The rotation measure maps of 7C1813+684, 3C65 and 3C268.1
contained obvious jumps in position angle which we were not able to
remove. Plots analogous to Figures \ref{4lambda} to \ref{6lambda}
indicated that there were regions that obviously contained errors
caused by n$\pi$ ambiguities. As previously noted the A array AD429
data was problematic and this was found to be the cause of the
jumps. To overcome this problem we shifted the position angles at 1.4
GHz data down by 10 to 15 degrees when the PA maps were
produced. This resolved any ambiguities.

Table \ref{vlaobsc} shows that all sources in sample C were observed
with IFs separated by only 50 MHz or less at around 1.4 GHz. This
means that they were not well enough separated at 1.4 GHz to overcome
the $n\pi$ ambiguities.  To compensate for this lack of separation the
4.8 GHz observations were split into their two component frequencies,
4885 MHz and 4535 MHz. We then used 4 frequencies for the fit instead
of 3, but we are still only marginally sensitive to $n\pi$ jumps.  The
resulting rotation measure maps cover the same frequency range as
samples A and B but use different frequencies for the fit. This was
not possible in the case of 3C351 and 3C299, resulting in larger
uncertainties in the rotation measurements for these sources.  In
the case of Sample C any source that has a large range of rotation
measures ($> 80$ rad m$^{-2}$), the {\small AIPS} task {\small RM}
will force the rotation measure into a range $\pm 40$ rad m$^{-2}$
around the mean rotation measure. This is due to the lack of frequency
separation at 1.4 GHz and it can cause jumps. In the case of 3C457
these jumps were severe and we were unable to resolve them. The
rotation measure and magnetic field maps for this source were not
included in the analysis. The rotation measure varies smoothly over
all other sources in this sample. The error affects the absolute value of the rotation
measure for each source and therefore it does not
affect the difference in the rotation measure, dRM and the rms
variation in the rotation measure, $\sigma_{RM}$.

All rotation measure maps contain pixels only where polarised
flux was observed at levels above $3\sigma_{noise}$ at all three
frequencies. We chose the lower noise threshold to allow for a
larger coverage of RM measurements over the lobes. However, even
with this lower threshold there were still a few sources where there
was very little to no rotation measure observed.

All depolarisation, spectral index, rotation measure and magnetic
field direction maps were overlayed with contours of total intensity
at 4.8 GHz.

Tables \ref{A} to \ref{C} give the total flux, percentage polarised
flux both at 1.4 GHz and 4.8 GHz, the depolarisation, rotation measure
and spectral index, between 1.4 GHz and 4.8 GHz.  All properties are
averaged over source components like the source core or individual
lobes and given in the observers frame of reference. This averaging
reduces the effect of the noise features seen in the maps. The total
flux, polarised flux and rotation measure values are the averaged
values determined from the maps with the {\small AIPS} task {\small
TVSTAT}. Percentage polarisation, depolarisation and spectral index are
then calculated from these average values. All integrated
quantities in Tables \ref{A} to \ref{C} are calculated excluding
pixels below the $3\sigma_{noise}$ threshold.

\begin{table}
\hspace{0cm}
\caption{Details of the VLA observations for sample C} \label{vlaobsc}
\begin{tabular}{p{0.5cm}ccccp{0.5cm}}
Source & Array  & Frequency & Bandwidth &Observing  & Int.\\
& Config. & (MHz)&(MHz)& Program & (min)\\
3C16  & A & 1452,1502&25 & 14/09/87 (AL146) & 59\\
&  B & 1452,1502&25 & 25/11/87 (AL146) & 39\\
&  B & 4885,4535&50  & 20/05/01 (AD444)&20\\
&  & 4885,4535&50 & 17/11/87 (AH271)&10 \\
&  C & 4885,4535&50  & 12/06/00 (AD429) &20\\
3C42  & A & 1452,1502&25 & 14/09/87 (AL146)&40\\
&  B & 4885,4535&50  & 23/12/91 (AF213) &67\\
3C46  & A & 1452,1502&25 & 31/07/99 (AD429)&16 \\
&  B & 1452,1502&25 & 25/11/87 (AL146) &35\\
&  B & 4885,4535&50  & 20/05/01 (AD444) &20\\
&  C & 4885,4535&50  & 12/06/00 (AD429) &20\\
3C299  & A & 1452,1502&25 & 31/07/99 (AD429)&16 \\
&  B & 4835,4535&50  & 20/05/01 (AD444)&15\\
&  & 4885,4835&50 & 28/01/98 (AP331)&15 \\
3C341  & A & 1452,1502&25 & 14/09/87 (AL146)&38 \\
&  B & 1452,1502&25 & 25/11/87 (AL146) &47\\
&  B & 4885,4535&50  & 20/05/01 (AD444)&11\\
&  & 4935,4535&50 & 26/10/92 (AA133) &25\\
&  C & 4885,4535&50  & 12/06/00 (AD429)&20 \\
3C351 & A & 1452,1502&25 & 31/07/99 (AD429) &16\\
&  B & 1452,1502&25 & 25/11/87 (AL146)&56 \\
&  B & 4885,4835&50  & 19/07/86 (AB369)&12\\
&  & 4885,4535&50 & 20/5/01 (AD444)&16\\
&  C & 4885,4835&50  & 09/10/87 (AA64)&22 \\
3C457 & A & 1452,1502&25 & 31/07/99 (AD429)&16 \\
&  B & 1452,1502&25& 25/11/87 (AL146)&30 \\
&  B &4885,4535&50  & 20/05/01 (AD444) &50\\
&  C & 4885,4535&50  & 12/06/00 (AD429)&20\\
4C14.27  & A & 1452,1502&25 & 14/09/87 (AL146) &28\\
&  B & 4885,4535&50  & 20/05/01 (AD444)&17\\
\end{tabular}
\end{table}
\section{The Results}
\begin{figure}
\centerline{\psfig{file=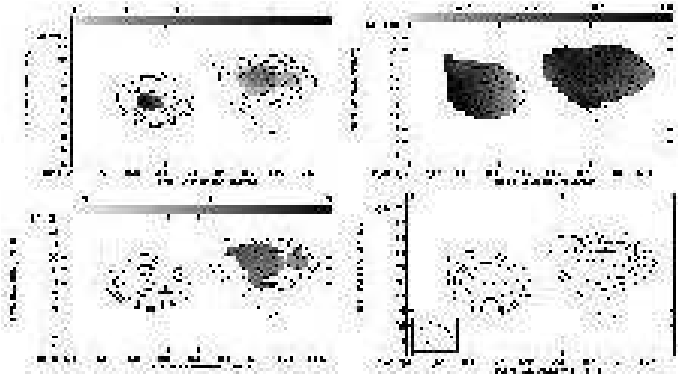,width=9cm}}
\caption{Maps of the radio source {\bf 6C0943+39}.(a - top left)
Map of the depolarisation between 4710 MHz and 1465 MHz. (b - top
right) Map of the spectral index between 4710 MHz and 1465 MHz. (c
- bottom left) Map of the rotation measure (rad m$^{-2}$) between
4710 MHz, 1665 MHz and 1465 MHz. (d - bottom right) Map of the
magnetic field direction (degrees).All contours are at $5 \sigma$
at 4710 MHz(0.5 mJy beam$^{-1}$)x (-1, 1, 2, 4...,1024). Beam size of 2.5'' x 1.4''.}
\label{0943}
\end{figure}
\begin{figure}
\centerline{\psfig{file=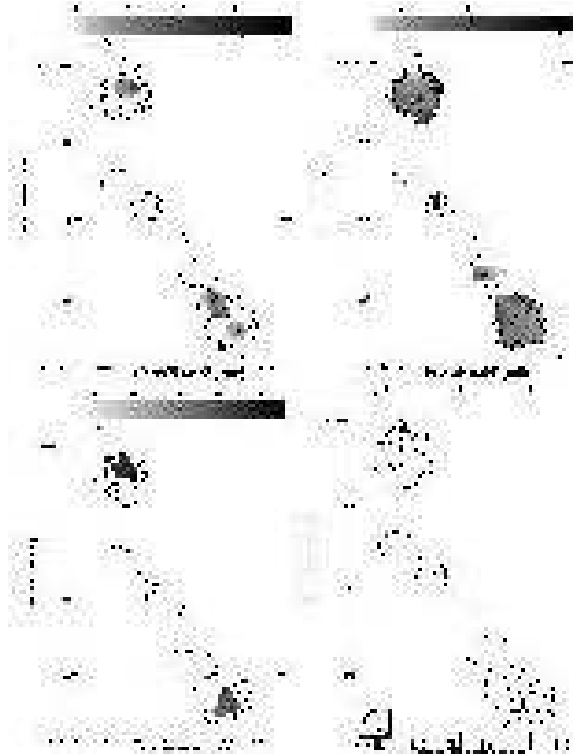,height=10cm,angle=0}} 
\caption{Maps of the radio source {\bf 6C1011+36}.(a
- upper left) Map of the depolarisation between 4535 MHz and 1465
MHz. (b - upper right) Map of the spectral index between 4535 MHz
and 1465 MHz. (c - lower right) Map of the magnetic field
direction (degrees). (d - lower left)Map of the rotation measure
(rad m$^{-2}$) between 4710 MHz, 1665 MHz and 1465 MHz. All contours
are at $5 \sigma$ at 4710 MHz(0.4 mJy beam$^{-1}$)x (-1, 1, 2,
4...,1024). Beam size of 3.5'' x 3.2''.} \label{1011}
\end{figure}
\begin{figure}
\centerline{\psfig{file=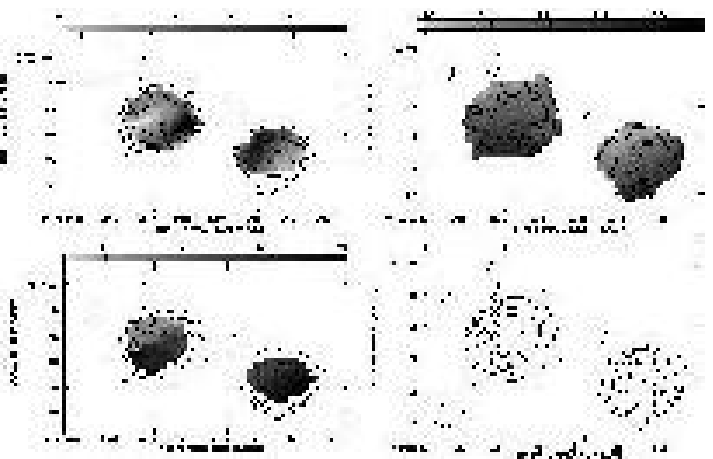,width=8cm}}
\caption{Maps of the radio source {\bf 3C324 }. (a - top left) Map
of the depolarisation between 4848 MHz and 1465 MHz. (b - top
right) Map of the spectral index between 4848 MHz and 1465 MHz. (c
- bottom left) Map of the rotation measure (rad m$^{-2}$) between
4848 MHz, 1665 MHz and 1465 MHz. (d - bottom right) Map of the
magnetic field direction (degrees). All contours are at $5 \sigma$
at 4848 MHz (1.5 mJy beam$^{-1}$)x (-1, 1, 2, 4...,1024). Beam size of
2.2'' x 1.6''.}
\label{3c324}
\end{figure}

In Figures \ref{0943} to \ref{4c14}, maps of the radio properties
discussed above are presented for all the sources from the 3
samples. Each figure shows the depolarisation map (where the
polarisation was detected at all frequencies), the spectral index map,
the rotation measure map and the magnetic field direction map (when a
rotation measure is detected), if no previously published maps
exist. For the three 7C sources, 6C1018+37, 3C65, 3C267 and 3C46
polarisation maps have also been included for both 1.4 GHz and 4.8 GHz as
there are no published polarisation maps of these sources. Table
\ref{maps} contains a listing of published maps for all sources. In all cases
only regions from the top end of the grey-
scales saturate, as we have always kept the lowest values well inside the
grey-scales, to ensure that no information has been lost.

\begin{table}
\caption{Details of previously published maps. TI = total intensity, P = polarisation, S = spectral index \& D = depolarisation} \label{maps}
\begin{tabular}{cccc}
Source & Map&Freq. & Ref.\\
& & (GHz) \\
6C0943+39& P&4.8 & \citet{bel99}\\
6C1011+36&P&4.8 & \citet{bel99}\\
&TI&1.4 & \citet{lla95}\\
6C1129+37&P&4.8 & \citet{bel99}\\
&TI&1.4 & \citet{lla95}\\
6C1256+36&P&4.8 & \citet{bel99}\\
&TI&1.4 & \citet{lla95}\\
6C1257+36&P&4.8 & \citet{bel99}\\
&TI&1.4 & \citet{lla95}\\
3C65& TI& 1.4, 4.8 & \citet{pwx95}\\
3C68.1 &  P &  4.8 & \citet{bhl94}\\
&TI &1.4 & \citet{lms89}\\
3C252& P &4.8 & \citet{fbb93}\\
3C265& P & 4.8 & \citet{fbb93}\\
3C267 & TI & 4.8 & \citet{blr97}\\
& & 1.4 & \citet{lms89}\\
3C268.1 & P & 4.8 & \citet{l81}\\
&TI &1.4 & \citet{lms89}\\
3C280 & P & 1.4, 4.8 & \citet{lp91}\\
& S, D & 1.4, 4.8 & \citet{lp91}\\
3C324 & TI & 4.8 & \citet{blr97}\\
&P & 1.4 & \citet{fbb93}\\
4C16.49 & P & 4.8 & \citet{lbm93}\\
3C16 & TI & 4.8 & \citet{gfg88}\\
&P & 1.4 & \citet{lp91a}\\
3C42 & P & 4.8 & \citet{fbp97}\\
3C46 & TI  & 4.8 & \citet{gfg88}\\
& & 1.4 & \citet{gpp88}\\
3C299 & P & 1.4, 4.8 & \citet{lp91}\\
 & S, D & 1.4, 4.8 & \citet{lp91b}\\
3C341 & P & 1.4 & \citet{lp91a}\\
3C351& TI&  4.8 & \citet{bhl94}\\
& P & 1.4 & \citet{lp91a}\\
3C457& P & 1.4&\citet{lp91a}\\
4C14.27 & P & 1.4 & \citet{lp91a}\\
\end{tabular}
\end{table}

In order to check the quality of our maps we also used the
maximum entropy method implemented in the AIPS task VTESS instead of
the CLEAN algorithm. The resulting maps are not significantly
different from those produced by the CLEAN algorithm. In fact, VTESS
is not necessarily superior to CLEAN in producing accurate maps of
extended low surface brightness regions \citep{rupen97}.

Depolarisation shadows (regions where the depolarisation is
appreciably greater than in the surrounding area) are seen in a few
sources, e.g. Figure \ref{3c324}, and may be real features.  These were
first found by \citet{feb89} in a study of Fornax A. 
Depolarisation shadows can be caused by the parent
galaxy as in the case of 3C324 or by an external galaxy in the
foreground of the source, causing a depolarising silhouette
\citep{blr972}. 

Polarisation angle measurements are ambiguous by $\pm$n$\pi$ and this can introduce 
ambiguities in the rotation measure. A change of $\pi$ between 
1.4 GHz and 5 GHz introduced by the fitting algorithm will cause a 
change of $\approx$ 80 rad m$^{-2}$ in rotation measure.
To determine if any strong rotation measure feature
is real a plot of the polarisation angle against $\lambda^2$, 
including $n \pi$ ambiguities can be produced. The best fit from {\small AIPS}
is then overlayed. Any true feature will not show any errors in $n
\pi$. This has been done for two sources: 4C16.49 and
6C1256+36. The resulting fits are presented in Figures \ref{4lambda} to \ref{6lambda} and
will be discussed in the relevant notes on these sources. The corresponding $\chi^2$ 
values for the fits are tabulated in Table \ref{values}. Another test that a feature 
is real is that a true jump in rotation measure causes depolarisation near
the jump but the magnetic field map shows no corresponding jump in the
region.

\subsection{Notes on the individual sources:}
\subsubsection{Sample A:}
\begin{figure}
\centerline{\psfig{file=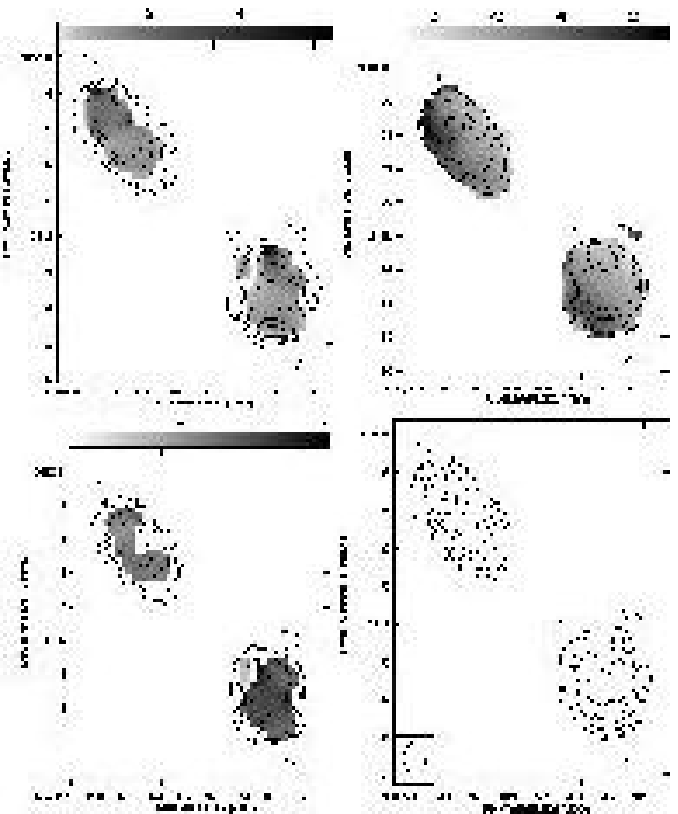,width=8cm,angle=0}}
\caption{Maps of the radio source {\bf 6C1256+36}. (a - upper
left) Map of the depolarisation between 4710 MHz and 1465 MHz. (b
- upper right) Map of the spectral index between 4710 MHz and 1465
MHz. (c - lower right) Map of the magnetic field
direction (degrees). (d - lower left)Map of the rotation measure
(rad m$^{-2}$) between 4710 MHz, 1665 MHz and 1465 MHz. All contours
are at $5 \sigma$ at 4710 MHz (0.25mJy beam$^{-1}$)x (-1, 1, 2,
4...,1024). Beam size of 2.5'' x 1.4''.} \label{1256}
\end{figure}

\begin{figure}
\centerline{\psfig{file=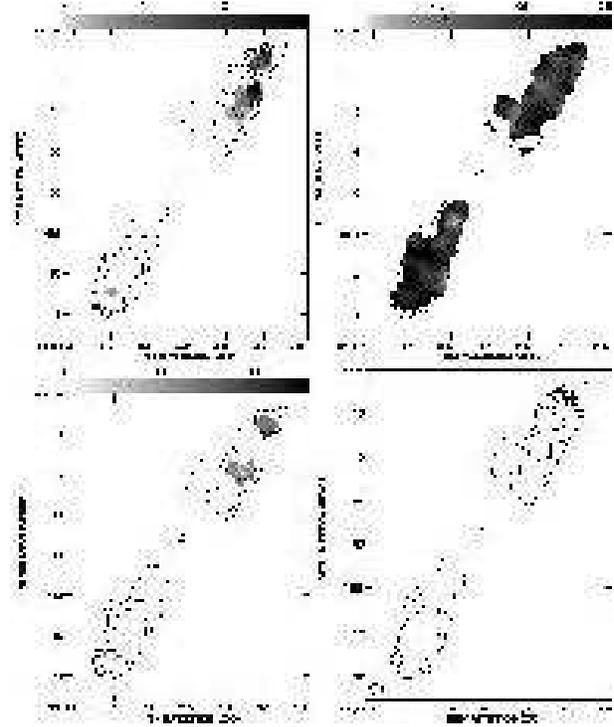,width=8cm,angle=0}}
\caption{Maps of the radio source {\bf 6C1257+37}.(a - upper left)
Map of the depolarisation between 4710 MHz and 1465 MHz. (b -
upper right) Map of the spectral index between 4860 MHz and 1465
MHz. (c - lower right) Map of the magnetic field direction
(degrees). (d - lower left) Map of the rotation measure (rad
m$^{-2}$) between 4710 MHz, 1665 MHz and 1465 MHz. All contours are
at $5 \sigma$ at 4710 MHz (0.25 mJy beam$^{-1}$)x (-1, 1, 2,
4...,1024). Beam size of 2.0'' x 1.4''.} \label{1257}
\end{figure}

\begin{figure}
\centerline{\psfig{file=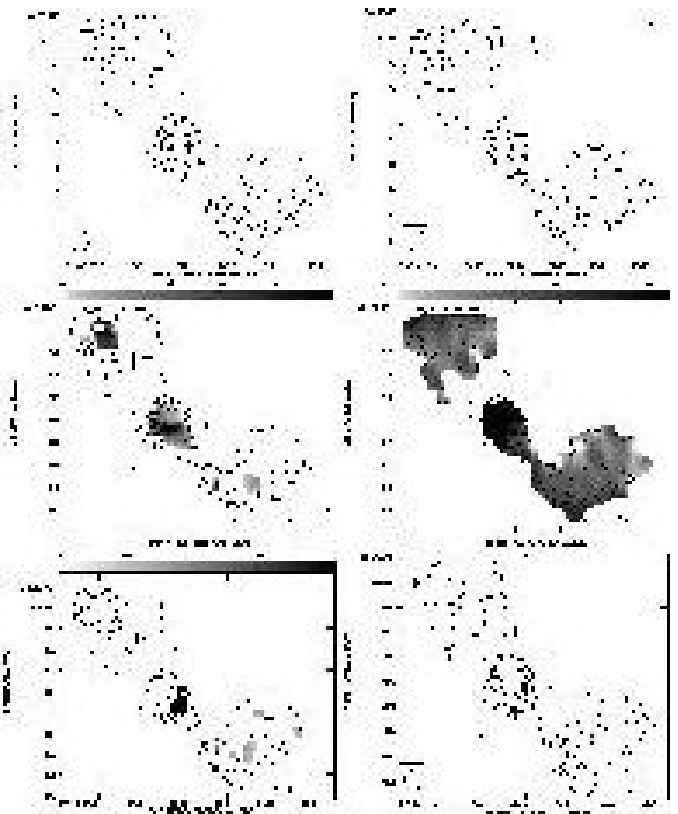,height=10cm,angle=0}}
\caption{Maps of the radio source {\bf 7C1745+642}. (a - upper
left) 4710 MHz total intensity map with vectors of polarisation
overlaid. The contour levels are at $5 \sigma$ (0.3mJy
beam$^{-1}$)x (-1, 1, 2, 4...,1024). 1 arc second corresponds to
$1.7\times10^{-3}{\rm Jy beam}^{-1}$. (b - upper right) 1465 MHz
total intensity map with vectors of polarisation overlaid. The
contour levels are at $3 \sigma$ (0.6 mJy beam$^{-1}$)x (-1, 1,
2,, 4...,1024). 1 arc second corresponds to $1.7\times10^{-3}{\rm
Jy beam}^{-1}$  (c - middle left) Map of the depolarisation
between 4710 MHz and 1465 MHz. (d - middle right) Map of the
Spectral Index between 4710 MHz and 1465 MHz. (e - lower right)
Map of the magnetic field direction (degrees). (f - lower left)
Map of the rotation measure (rad m$^{-2}$) between 4710 MHz, 1665 MHz
and 1465 MHz. All contours (c-f) are at $5 \sigma$ at 4710 MHz
(0.3 mJy beam$^{-1}$)x (-1, 1, 2, 4...,1024). Beam size of 1.7'' x 1.2''.} \label{1745}
\end{figure}

\begin{figure}
\centerline{\psfig{file=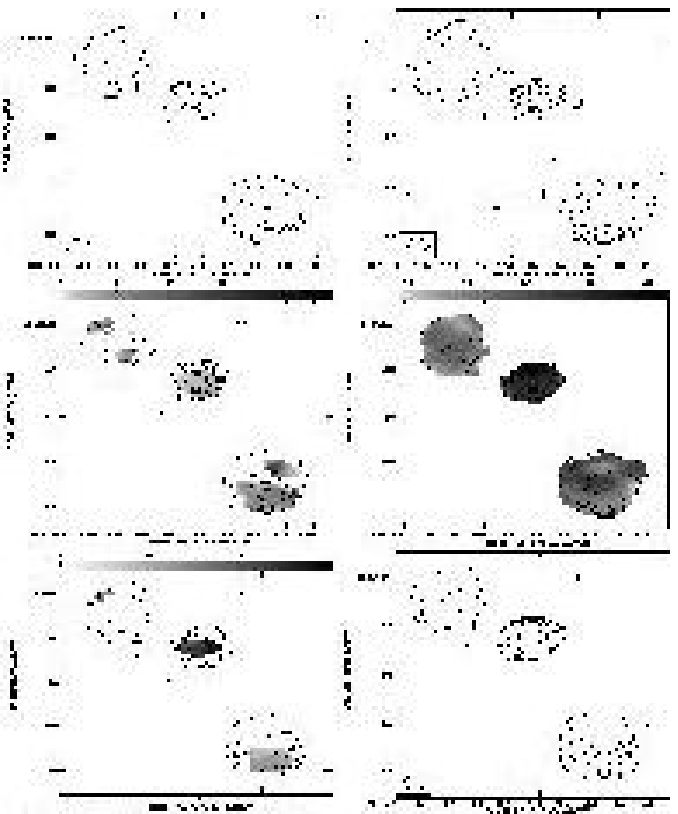,height=10cm,angle=0}}
\caption{Maps of the radio source {\bf 7C1801+690}. (a - upper
left) 4710 MHz total intensity map with vectors of polarisation
overlaid. The contour levels are at $5 \sigma$ (0.3 mJy
beam$^{-1}$)x (-1, 1, 2, 4..., 1024). 1 arc second corresponds to
$1.7\times10^{-3}{\rm Jy beam}^{-1}$. (b - upper right) 1465 MHz
total intensity map with vectors of polarisation overlaid. The
contour levels are at $3 \sigma$ (0.6 mJy beam$^{-1}$)x (-1, 1, 2,
4,..,1024). 1 arc second corresponds to $1.7\times10^{-3}{\rm Jy
beam}^{-1}$. (c - middle left) Map of the depolarisation between
4710 MHz and 1465 MHz. (d - middle right) Map of the spectral index
between 4710 MHz and 1465 MHz. (e - lower right) Map of the magnetic
field direction (degrees). (f - lower left) Map of the rotation
measure (rad m$^{-2}$) between 4710 MHz, 1665 MHz and 1465 MHz. All
contours (c-f) are at $5 \sigma$ at 4710 MHz (0.30 mJy
beam$^{-1}$)x (-1, 1, 2, 4...,1024). Beam size of 2.2'' x 1.5''.} \label{1801}
\end{figure}

\begin{figure}
\centerline{\psfig{file=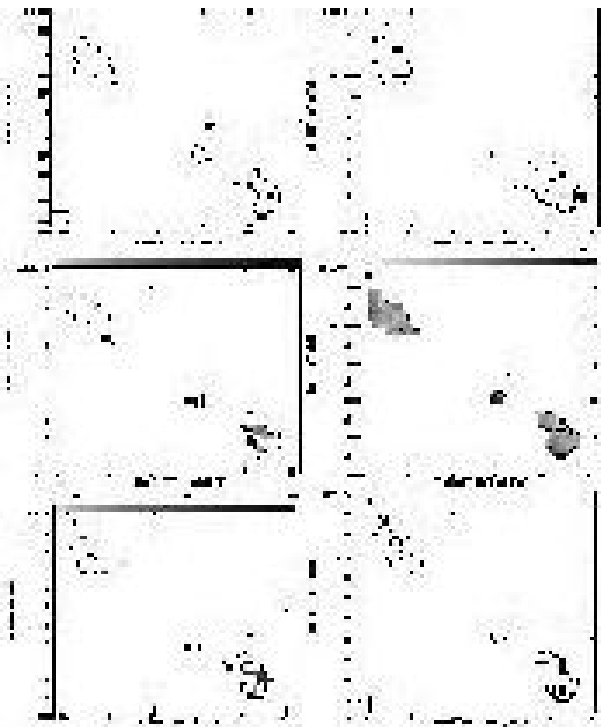,height=10cm,angle=0}}
\caption{Maps of the radio source {\bf 7C1813+684}. (a - upper
left) 4710 MHz total intensity map with vectors of polarisation
overlaid. The contour levels are at $5 \sigma$ (0.35 mJy
beam$^{-1}$)x (-1, 1, 2, 4... ,1024). 1 arc second corresponds to
$8.3\times10^{-4}{\rm Jy beam}^{-1}$. (b - upper right) 1465 MHz
total intensity map with vectors of polarisation overlaid. The
contour levels are at $3 \sigma$ (0.7 mJy beam$^{-1}$)x (-1, 1, 2,
4... ,1024). 1 arc second corresponds to $1.7\times10^{-3}{\rm Jy
beam}^{-1}$. (c - middle left) Map of the depolarisation between
4710 MHz and 1465 MHz. (d - middle right) Map of the spectral
index between 4710 MHz and 1465 MHz. (e - lower right) Map of the
magnetic field direction (degrees). (f - lower left) Map of the
rotation measure (rad m$^{-2}$) between 4710 MHz, 1665 MHz and 1465
MHz. All contours (c-f) are at $5 \sigma$ at 4710 MHz (0.35 mJy
beam$^{-1}$)x (-1, 1, 2, 4...,1024). Beam size of 2.5'' x 2.0''.} \label{1813}
\end{figure}

\paragraph*{{\it 6C0943+39:}}(Figure \ref{0943}) No core is detected in the observations, \citet{bel99}
detected a core at 8.2 GHz and minimally at 4.8 GHz. Our non-detection is probably due to the different 
resolution of the data. The value of the rotation measure in the Eastern lobe
must be considered with caution as it is based on only a few
pixels.
\paragraph*{{\it 6C1011+36:}}(Figure \ref{1011}) This is a classic double-lobed structure, 
showing a strong core at both 4.7 GHz and 1.4 GHz with an inverted spectrum.
\begin{figure}
\centerline{\psfig{file=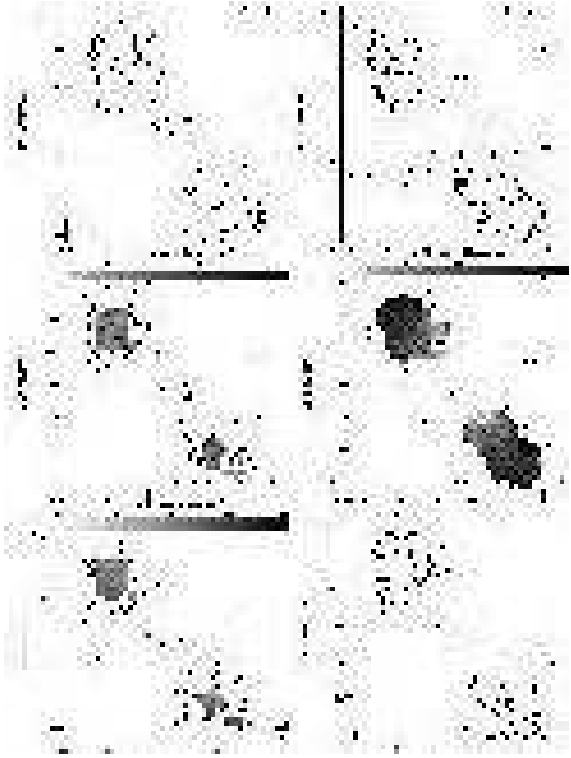,height=10cm,angle=0}}
\caption{Maps of the radio source {\bf 6C1018+37}. (a - upper
left) 4710 MHz total intensity map with vectors of polarisation
overlaid. The contour levels are at $5 \sigma$ (100 $\mu${\rm Jy
beam}$^{-1}$)x (-1, 1,2, 4...,1024). 1 arc second corresponds to
$1.7\times10^{-3}{\rm Jy beam}^{-1}$. (b - upper right) 1465 MHz
total intensity map with vectors of polarisation overlaid. The
contour levels are at $3 \sigma$ (0.5 mJy beam$^{-1}$)x (-1, 1,2,
4,...,1024). 1 arc second corresponds to $1.7\times10^{-3}{\rm Jy
beam}^{-1}$ (c - middle left) Map of the depolarisation between
4710 MHz and 1465 MHz. (d - middle right) Map of the Spectral
Index between 4710 MHz and 1465 MHz. (e - lower right) Map of the
magnetic field direction (degrees). (f - lower left) Map of the
rotation measure (rad m$^{-2}$) between 4710 MHz, 1665 MHz and 1465
MHz. All contours (c-f) are at $5 \sigma$ at 4885 MHz (100 $\mu$Jy
beam$^{-1}$)x (-1, 1, 2, 4...,1024). Beam size of 4'' x 4''.} \label{1018}
\end{figure}
\paragraph*{{\it 6C1018+37:}}(Figure \ref{1018}) 
The maps were made with the smaller arrays only at each frequency. In the
1.4 GHz A-array data set the lower lobe was partially resolved out but this was compounded 
by the high noise, so
no feasible combination of the A and B array was possible.
To maintain consistency the B-array 4.7 GHz data was also excluded.
\begin{figure}
\centerline{\psfig{file=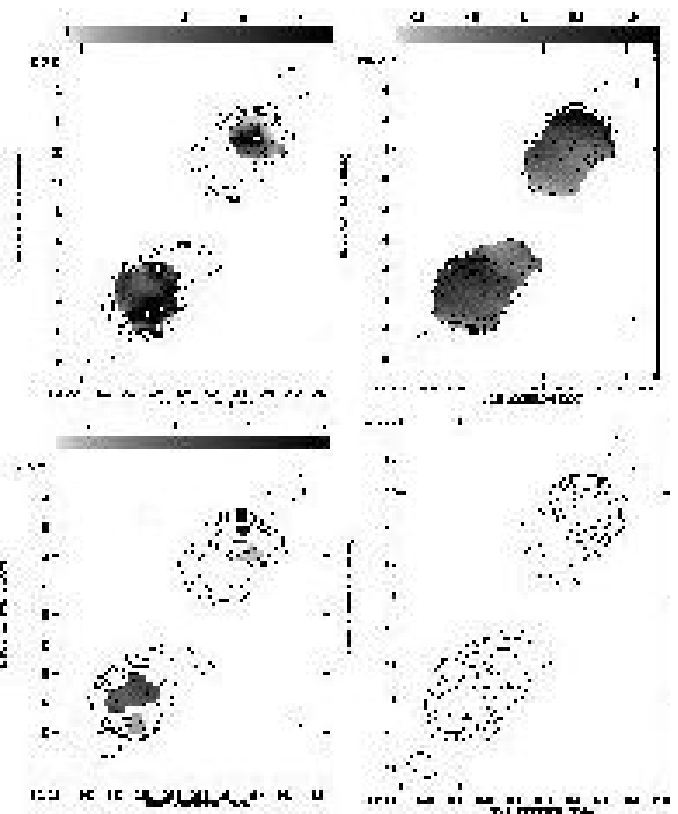,width=8cm,angle=0}}
\caption{Maps of the radio source {\bf 6C1129+37}. (a - upper
left) Map of the depolarisation between 4710 MHz and 1465 MHz. (b
- upper right) Map of the spectral index between 4710 MHz and 1465
MHz. (c - lower right) Map of the magnetic field direction
(degrees). (d - lower left) Map of the rotation measure (rad
m$^{-2}$) between 4710 MHz, 1665 MHz and 1465 MHz. All contours are
at $5 \sigma$ at 4710 MHz(0.3mJy beam$^{-1}$)x (-1, 1, 2,
4...,1024). Beam size of 2.5'' x 1.4''.} \label{1129}
\end{figure}
\paragraph*{{\it 6C1129+37:}}(Figure \ref{1129}) 
The SE lobe contains two distinct hotspots.
\citet{bel99} found 3 hotspots. Our non-detection is probably due to the 
different resolutions of the two observations. The source shows distinct regions of very strong
depolarisations, however these regions are slightly smaller than
the beam size. 
\paragraph*{{\it 6C1256+36:}}(Figure \ref{1256}) 
The rotation measure map
shows distinct changes in the values of the rotation measure. Plot \ref{6lambda}
shows that although the jump in RM does not correspond to a jump in depolarisation
it is not due to any error in the fitting program.
\paragraph*{{\it 6C1257+37:}}(Figure \ref{1257})
A core was detected at 4.7 GHz but was absent from the
1.4 GHz data. The high noise
level and short observation time meant that the S lobe had very
little polarised flux above the noise level, resulting in a reliable
value for the rotation measure being found in only a few pixels
around the hotspots.
\paragraph*{{\it 7C1745+642:}}(Figure \ref{1745}) This is a highly core dominated source,
with the northern lobe appearing faintly. There is an
indication of a jet-like structure leading down from the core into
the southern, highly extended, off-axis, lobe. The source is a
weak core dominated quasar \citep{bh01}.
\paragraph*{{\it 7C1801+690:}}(Figure \ref{1801}) This is an asymmetric core
dominated quasar \citep{bh01}. The N lobe is very faint
and appears to be much closer to the core component than the more
extended S lobe. It shows very little polarisation compared to the
relatively strong polarisation of the core and S components. 
\paragraph*{{\it 7C1813+684:}}(Figure \ref{1813}) This is the faintest
of the sources in sample A and is also a quasar \citep{bh01}.
The source shows a compact core that is present at all
observing frequencies, but it is too faint to detect any
reliable polarisation properties.

\subsubsection{Sample B:}
\begin{figure}
\centerline{\psfig{file=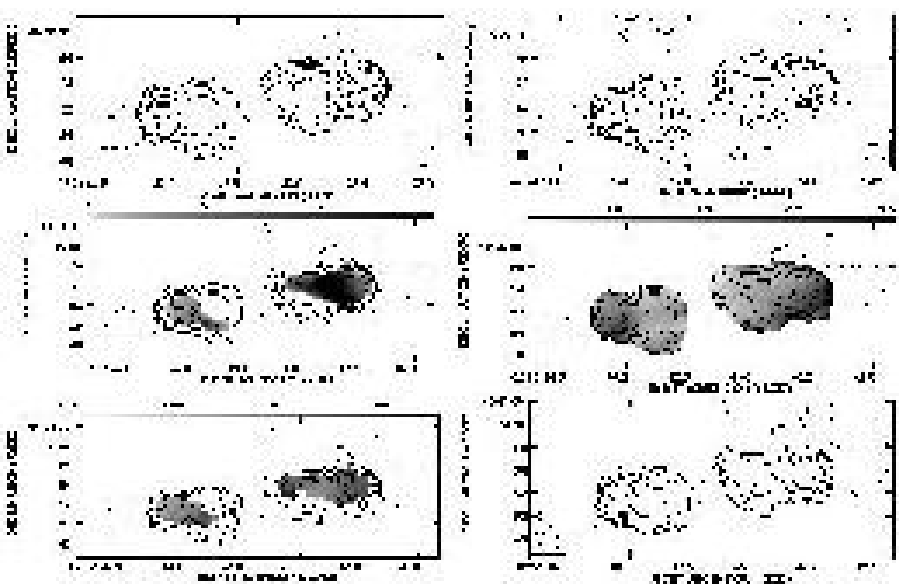,width=10cm}}
\caption{Maps of the radio source {\bf 3C65}.  (a - upper left)
4710 MHz total intensity map with vectors of polarisation
overlaid. The contour levels are at $5 \sigma$ (0.8 mJy
beam$^{-1}$)x (-1, 1, 2, 4... ,1024). 1 arc second corresponds to
$1.7\times10^{-2}{\rm Jy beam}^{-1}$. (b - upper right) 1465 MHz
total intensity map with vectors of polarisation overlaid. The
contour levels are at $3 \sigma$ (3.0 mJy beam$^{-1}$)x (-1, 1, 2,
4... ,1024). 1 arc second corresponds to $8.3\times10^{-3}{\rm Jy
beam}^{-1}$. (c - middle left) Map of the depolarisation between
4710 MHz and 1465 MHz. (d - middle right) Map of the spectral
index between 4710 MHz and 1465 MHz. (e - bottom left) Map of the
rotation measure (rad m$^{-2}$) between 4710 MHz, 1665 MHz and 1465
MHz. (f - bottom right) Map of the magnetic field direction
(degrees). All contours (c-f) are at $5 \sigma$ at 4860 MHz(0.8
mJy beam$^{-1}$)x (-1, 1, 2, 4...,1024). Beam size of 2.0'' x 1.5''.} \label{3c65}
\end{figure}

\begin{figure}
\centerline{\psfig{file=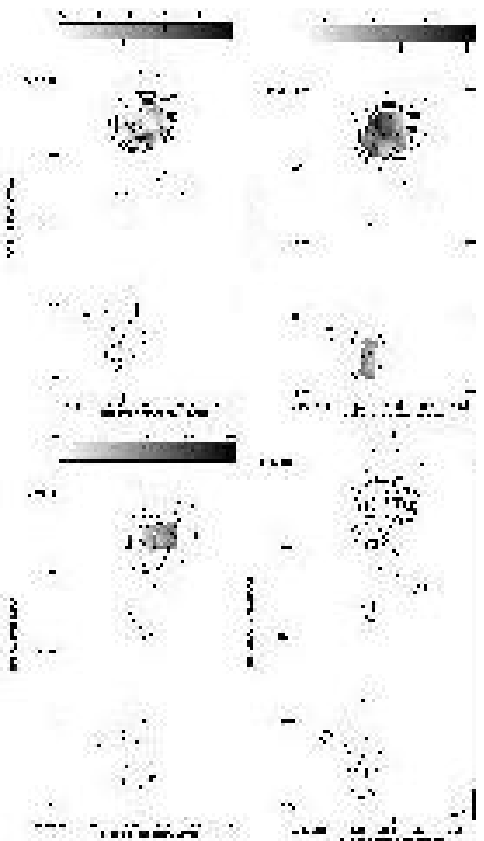,width=6cm,angle=0}}
\caption{Maps of the radio source {\bf 3C68.1}. (a - upper left)
Map of the depolarisation between 4710 MHz and 1417 MHz. (b -
upper right) Map of the spectral index between 4710 MHz and 1417
MHz. (c - lower right) Map of the magnetic field direction
(degrees). (d - lower left) Map of the rotation measure (rad
m$^{-2}$) between 4710 MHz, 1662 MHz and 1417 MHz. All contours are
at $5 \sigma$ at 4710 MHz (2.0 mJy beam$^{-1}$)x (-1, 1, 2,
4...,1024). Beam size of 3.5'' x 3.5''.} \label{3c68}
\end{figure}

\begin{figure}
\centerline{\psfig{file=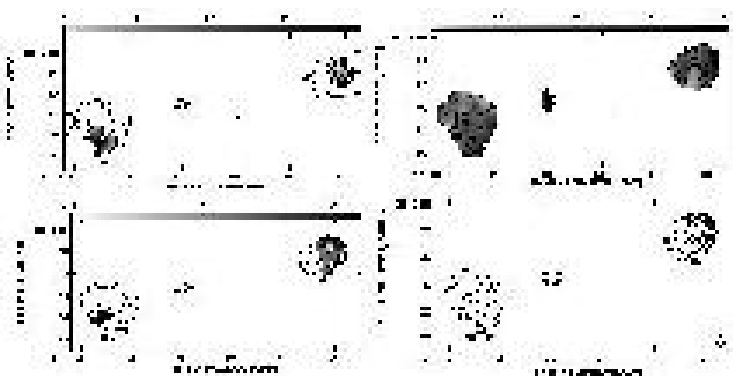,width=8cm}}
\caption{Maps of the radio source {\bf 3C252 }. (a - top left) Map
of the depolarisation between 4710 MHz and 1465 MHz. (b - top
right) Map of the spectral index between 4710 MHz and 1465 MHz. (c
- bottom left) Map of the rotation measure (rad m$^{-2}$) between
4710 MHz, 1665 MHz and 1465 MHz. (d - bottom right) Map of the
magnetic field direction (degrees). All contours are at $5 \sigma$
at 4710 MHz (0.4 mJy beam$^{-1}$)x (-1, 1, 2, 4...,1024). Beam size of
2.0'' x 2.5''.}
\label{3c252}
\end{figure}

\begin{figure}
\centerline{\psfig{file=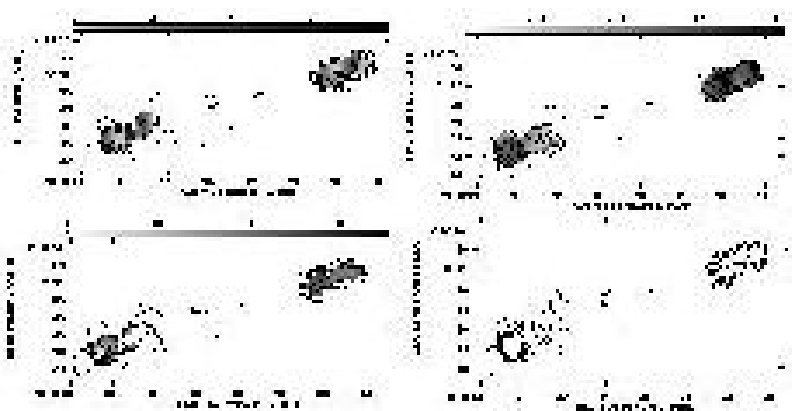,width=8cm}}
\caption{Maps of the radio source {\bf 3C265}.(a - top left) Map of the
depolarisation between 4848 MHz and 1417 MHz. (b - top right)
Map of the spectral index between 4848 MHz and 1417 MHz.
(c - bottom left) Map of the rotation measure (rad
m$^{-2}$) between 4848 MHz, 1662 MHz and 1417 MHz. (d - bottom right) Map
of the magnetic field direction (degrees) All contours are at $5
\sigma$ at 4848 MHz (0.7 mJy beam$^{-1}$)x (-1, 1, 2, 4...,1024). Beam
size of 3.5'' x 3.0''.}
\label{3c265}
\end{figure}

\begin{figure}
\centerline{\psfig{file=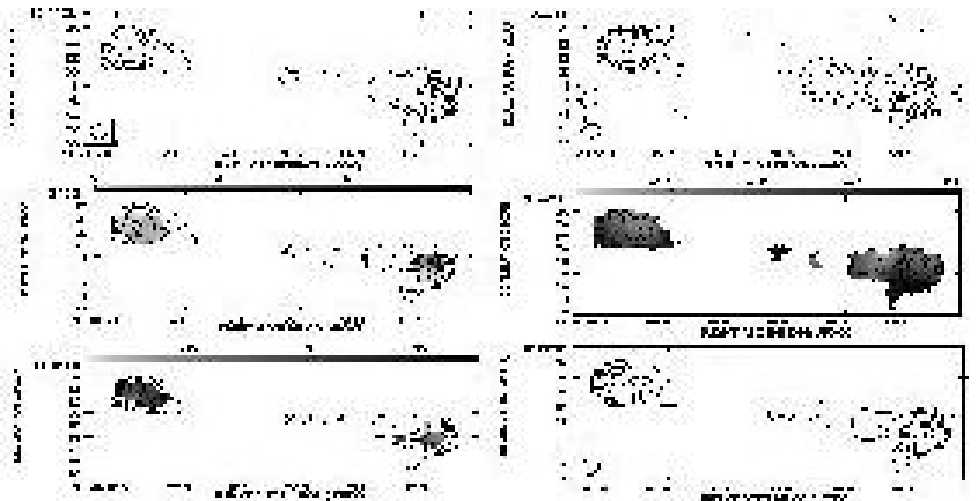,width=8cm}}
\caption{Maps of the radio source {\bf 3C267 }.  (a - upper left)
4848 MHz total intensity map with vectors of polarisation
overlaid. The contour levels are at $5 \sigma$ (0.8 mJy
beam$^{-1}$)x (-1, 1, 2, 4... ,1024). 1 arc second corresponds to
$8.3\times10^{-3}{\rm Jy beam}^{-1}$. (b - upper right) 1465 MHz
total intensity map with vectors of polarisation overlaid. The
contour levels are at $3 \sigma$ (1.3 mJy beam$^{-1}$)x (-1, 1, 2,
4... ,1024). 1 arc second corresponds to $8.3\times10^{-3}{\rm Jy
beam}^{-1}$. (c - middle left) Map of the depolarisation between
4848 MHz and 1465 MHz. (d - middle right) Map of the spectral
index between 4848 MHz and 1465 MHz. (e - bottom left) Map of the
rotation measure (rad m$^{-2}$) between 4848 MHz, 1665 MHz and 1465
MHz. (f - bottom right) Map of the magnetic field direction
(degrees). All contours (c-f)are at $5 \sigma$ at 4848 MHz (0.8
mJy beam$^{-1}$)x (-1, 1, 2, 4...,1024). Beam size of 2.2'' x 2.0''.} \label{3c267}
\end{figure}

\begin{figure}
\centerline{\psfig{file=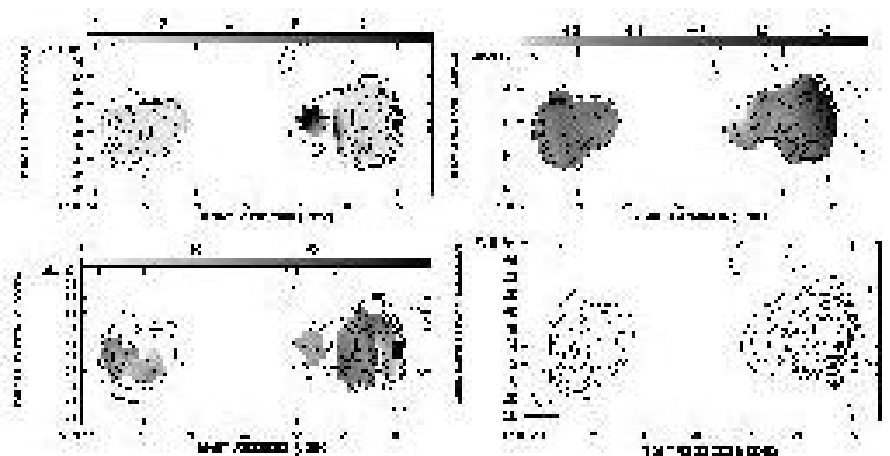,width=8cm}}
\caption{Maps of the radio source {\bf 3C268.1}.(a - top left) Map of
the depolarisation between 4848 MHz and 1417 MHz. (b - top
right) Map of the spectral index between 4848 MHz and 1417 MHz.
(c - bottom left) Map of the rotation measure (rad m$^{-2}$) between
4848 MHz, 1662 MHz and 1417 MHz. (d - bottom right) Map of the magnetic
field direction (degrees). All contours are at $5 \sigma$ (5.2 mJy
beam$^{-1}$)x (-1, 1, 2, 4...,1024). Beam size of 4.0'' x 3.5''.} \label{3c268.1}
\end{figure}

\begin{figure}
\centerline{\psfig{file=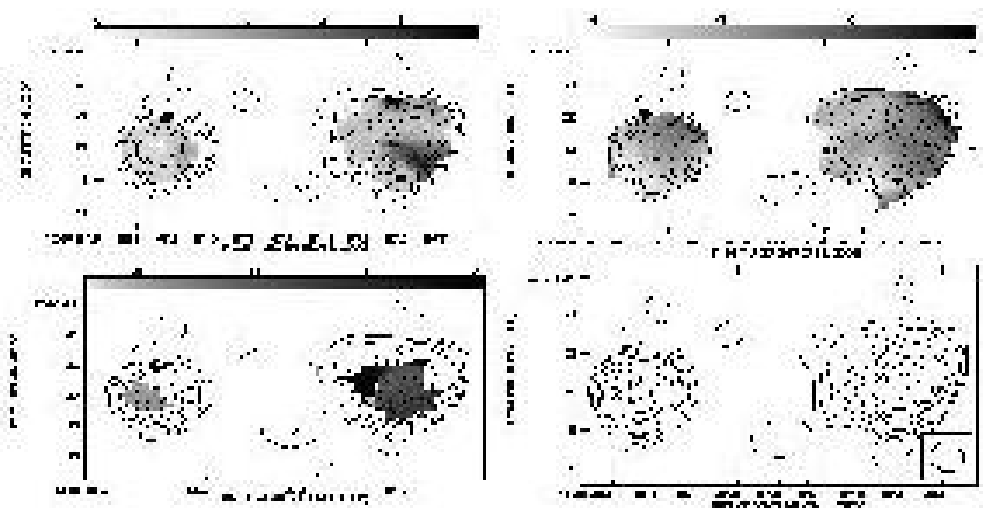,width=8cm,angle=0}}
\caption{Maps of the radio source {\bf 3C280 }. 
(a - top left) Map of the depolarisation between 4848 MHz and 1465 MHz.
(b - top right) Map of the spectral index between 4848 MHz and 1465 MHz.
(c - bottom left) Map of
the rotation measure (rad m$^{-2}$) between 4848 MHz, 1665 MHz and
1465 MHz. (d - bottom right) Map of the magnetic field direction
(degrees). All contours are at $5 \sigma$ at 4848 MHz (0.8 mJy
beam$^{-1}$)x (-1, 1, 2, 4...,1024). Beam size of 1.6'' x 1.6''.} \label{3c280}
\end{figure}

\begin{figure}
\centerline{\psfig{file=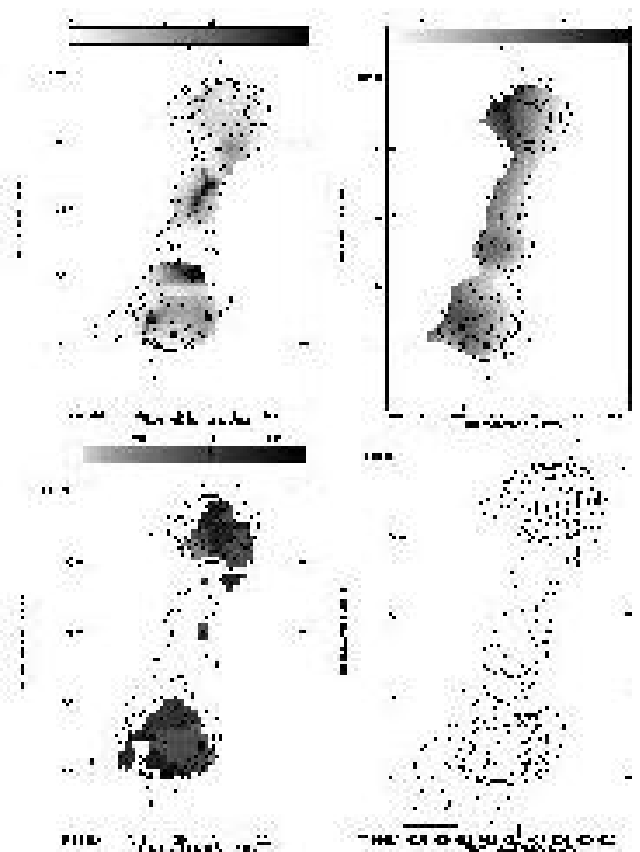,width=8cm,angle=0}}
\caption{Maps of the radio source {\bf 4C16.49}. (a - upper left)
Map of the depolarisation between 4710 MHz and 1465 MHz. (b -
upper right) Map of the spectral index between 4710 MHz and 1465
MHz. (c - lower right) Map of the magnetic field direction
(degrees). (d - lower left) Map of the rotation measure (rad
m$^{-2}$) between 4710 MHz, 1665 MHz and 1465 MHz. All contours are
at $5 \sigma$ at 4710 MHz (0.8mJy beam$^{-1}$)x (-1, 1, 2,
4...,1024). Beam size of 2.2'' x 1.8''.} \label{4c16}
\end{figure}

\begin{figure}
\centerline{\psfig{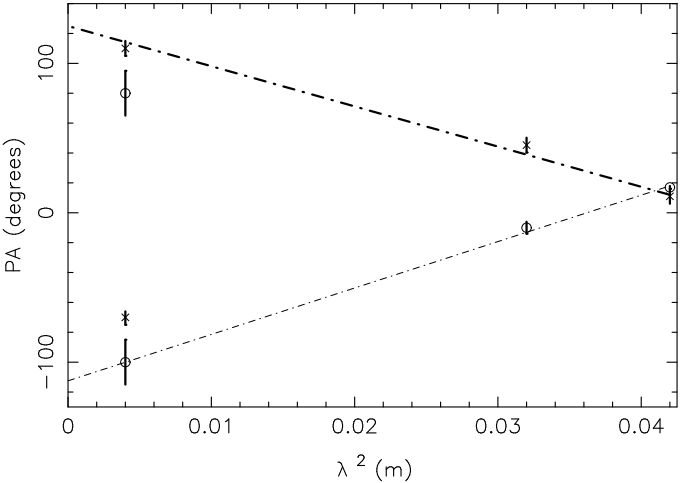}}
\caption{A plot of the polarisation angle against $\lambda^2$, allowing for n$\pi$, for the northern
lobe of 4C16.49 with the dashed lines showing the best fit models. All
n$\pi$ solutions for the 5 GHz data are considered and plotted. Data
for two small regions, one on each side of the jump are plotted with o
indicating one side of the jump and x the other. Table \ref{values}
show the $\chi^2$ values for each fit. The jump plotted lies SW of the central intensity contour.}\label{4lambda}
\end{figure}

\begin{figure}
\centerline{\psfig{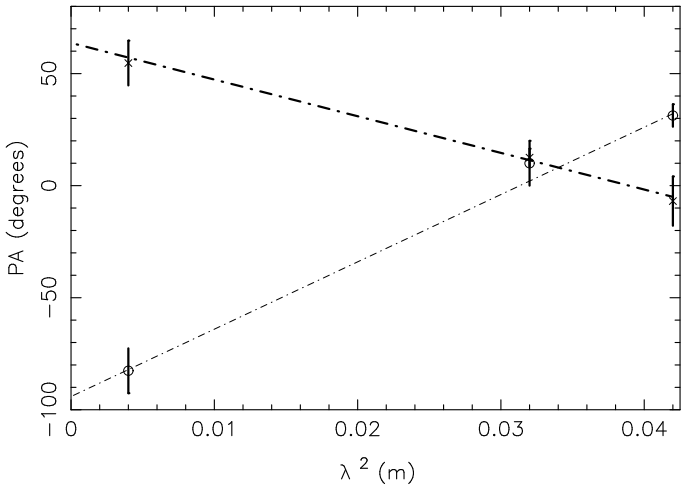}}
\caption{As Figure \ref{4lambda} but for 4C16.49 south lobe.Data
for two small regions, one on each side of the jump are plotted. These
regions lie either side of the jump south of the peak intensity
contour.}\label{42lambda}
\end{figure}

\begin{figure}
\centerline{\psfig{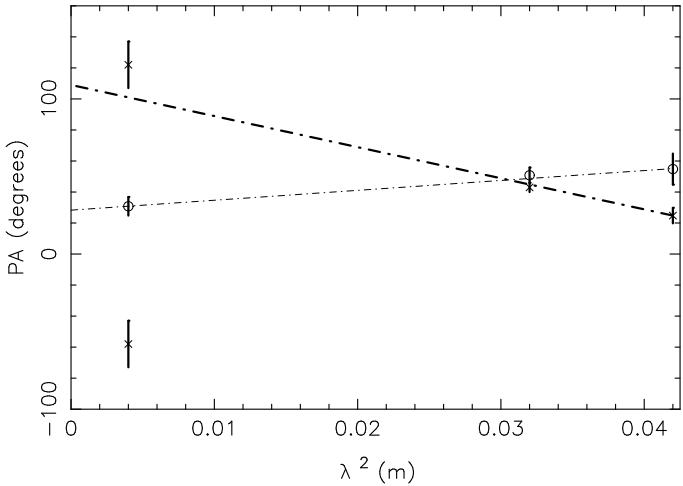}}
\caption{As Figure \ref{4lambda} but for 6C1256+36 south lobe.Data
for two small regions, one on each side of the jump are plotted.}\label{6lambda}
\end{figure}

\paragraph*{{\it 3C65:}}(Figure \ref{3c65})
The W lobe shows a strong depolarisation shadow that is smaller than the
beam size. \citet{b00} found the source to lie in a cluster which
might account for the presence of the depolarisation shadow and
the large depolarisation overall.
\paragraph*{{\it 3C68.1:}}(Figure \ref{3c68}) The source is a quasar \citep{bhl94}.
A core has been detected by \citet{bhl94} in deeper observations.
\paragraph*{{\it 3C252:}}(Figure \ref{3c252}) 
The SE lobe shows a sharp drop in the polarisation
between the 4.7 GHz and 1.4 GHz observations.
\paragraph*{{\it 3C265:}}(Figure \ref{3c265}) The NW
lobe shows evidence of a compact, bright region 
with a highly ordered magnetic field which \citet{fbb93} show is the primary
hotspot at higher resolutions.
\paragraph*{{\it 3C267:}}(Figure \ref{3c267}) The E lobe is highly extended, reaching to the core
position, which can be seen in the 1.4 GHz image. The large
depolarisation region in the W lobe coincides with a region of no
observed rotation measure. The core is strongly inverted with $\alpha
= 0.48$
\paragraph*{{\it 3C268.1:}}(Figure \ref{3c268.1}) The average spectral index over the entire source is
$\alpha=-0.66$ which is rather flat but taking the 4.8
 GHz data from \citet{gc91} and 1.4 GHz
data from \citet{lp80} the value is very similar, $\alpha=-0.68$. 
\paragraph*{{\it 3C280:}}(Figure \ref{3c280})
The value of the rotation measure and the magnetic field direction in the E
lobe must be treated with caution as it is based on only a small
region of the entire lobe. The sharp changes in the rotation
measure map are not seen in the magnetic field map and the depolarisation
map shows a similar structure suggesting 
that is not due a fitting error.
\paragraph*{{\it 3C324:}}(Figure \ref{3c324}) The NE lobe shows evidence of a
depolarisation shadow. \citet{b00} found the source to lie in a
cluster which may explain the faint shadow.
\paragraph*{{\it 4C16.49:}}(Figure \ref{4c16}) The source is a quasar \citep{bh01},
that shows a strong radio core, jet structure and possibly a small
counter-jet. The source is highly asymmetric with the lower lobe
almost appearing to connect to the core. It has a very steep
spectral index, $\alpha < -1.0$ making it an atypical source.
Figures \ref{4lambda} and \ref{42lambda} demonstrates that the sharp changes in the rotation
measure map are not due to any fitting errors.

\subsubsection{Sample C:}
\begin{figure}
\centerline{\psfig{file=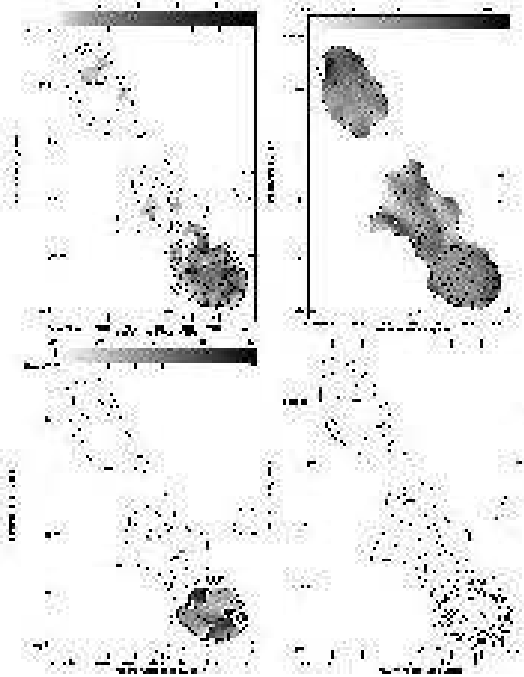,width=7cm,angle=0}}
\caption{Maps of the radio source {\bf 3C16}. (a - upper left) Map
of the depolarisation between 4710 MHz and 1452 MHz. (b - upper
right) Map of the spectral index between 4710 MHz and 1452 MHz. (c
- lower right) Map of the magnetic field direction (degrees). (d -
lower left) Map of the rotation measure (rad m$^{-2}$) between 4885
MHz, 4535 MHz, 1502 MHz and 1452 MHz. All contours are at $5 \sigma$ at 4710
MHz (0.40 mJy beam$^{-1}$)x (-1, 1, 2, 4...,1024). Beam size of 3.5'' x 3.5''.} \label{3c16}
\end{figure}

\begin{figure}
\centerline{\psfig{file=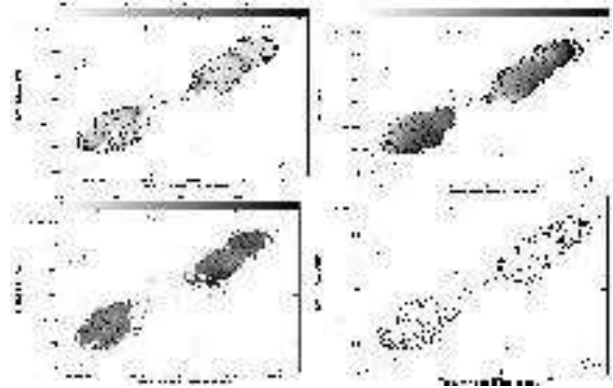,width=8cm}}
\caption{Maps of the radio source {\bf 3C42 }. (a - upper left)
Map of the depolarisation between 4710 MHz and 1452 MHz. (b -
upper right) Map of the spectral index between 4710 MHz and 1465
MHz. (c - lower right) Map of the magnetic field direction
(degrees). (d - lower left) Map of the rotation measure (rad
m$^{-2}$) between 4885 MHz, 4535 MHz, 1502 MHz and 1452 MHz. All contours are
at $5 \sigma$ at 4710 MHz (1.0 mJy beam$^{-1}$)x (-1, 1, 2,
4...,1024). Beam size of 2.5'' x 1.8''.} \label{3c42}
\end{figure}

\begin{figure}
\centerline{\psfig{file=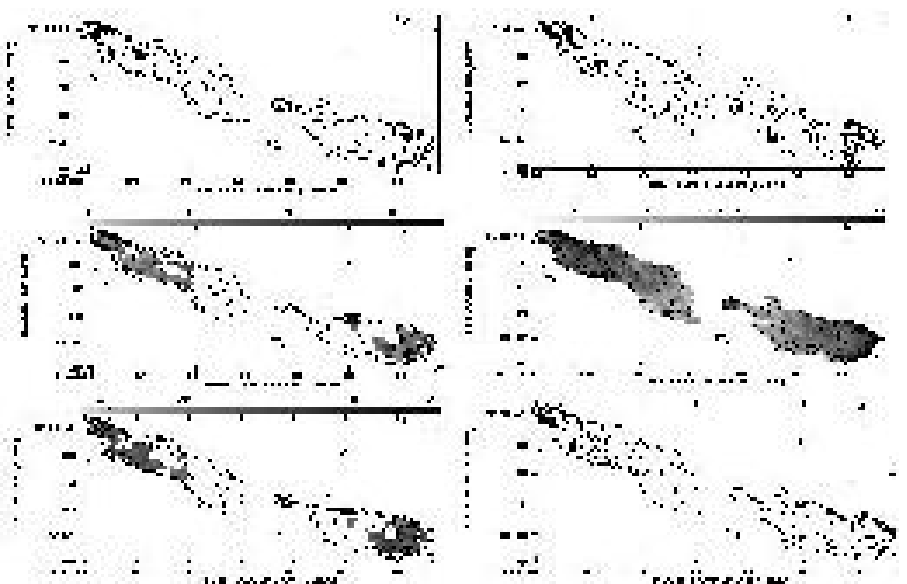,width=8cm}}
\caption{Maps of the radio source {\bf 3C46}.  (a - upper
left) 4710MHz total intensity map with vectors of polarisation
overlaid. The contour levels are at $5 \sigma$ (0.2 5mJy
beam$^{-1}$)x (-1, 1, 2, 4..., 1024). 1 arc second corresponds to
$1.7\times10^{-3}{\rm Jy beam}^{-1}$. (b - upper right) 1452MHz
total intensity map with vectors of polarisation overlaid. The
contour levels are at $3 \sigma$ (1.0 mJy beam$^{-1}$)x (-1, 1, 2,
4,..,1024). 1 arc second corresponds to $5.5\times10^{-4}{\rm Jy
beam}^{-1}$.(c - middle right) Map of the
depolarisation between 4710 MHz and 1452 MHz. 
(d - middle left) Map of the spectral index between 4710 MHz and 1452 MHz.
(e - bottom right) Map of
the rotation measure (rad m$^{-2}$) between 4885 MHz, 4535 MHz, 1502 MHz and
1452 MHz. (f - bottom left) Map of the magnetic field direction
(degrees). All contours (c- f) are at $5 \sigma$ at 4710 MHz (0.25mJy
beam$^{-1}$)x (-1, 1, 2, 4...,1024). Beam size of 4.5'' x 3.0''.} \label{3c46}
\end{figure}

\begin{figure}
\centerline{\psfig{file=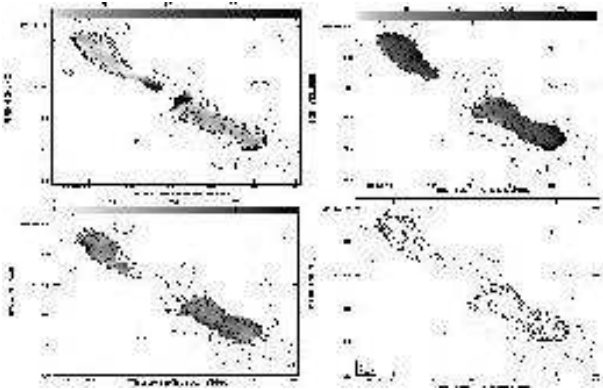,width=8cm}}
\caption{Maps of the radio source {\bf 3C341 }. (a - upper left)
Map of the depolarisation between 4710 MHz and 1452 MHz. (b -
upper right) Map of the spectral index between 4710 MHz and 1452
MHz. (c - lower right) Map of the magnetic field direction
(degrees). (d - lower left) Map of the rotation measure (rad
m$^{-2}$) between 4885 MHz, 4535 MHz, 1502 MHz and 1452 MHz. All contours  are
at $5 \sigma$ at 4710 MHz (0.7 mJy beam$^{-1}$)x (-1, 1, 2,
4...,1024). Beam size of 4.0'' x 4.0''.} \label{3c341}
\end{figure}

\begin{figure}
\centerline{\psfig{file=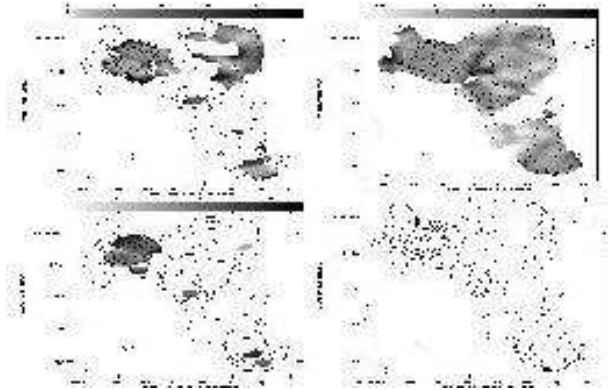,width=8cm}}
\caption{Maps of the radio source {\bf 3C351 }. (a - upper left)
Map of the depolarisation between 4810 MHz and 1452 MHz. (b -
upper right) Map of the spectral index between 4810 MHz and 1452
MHz. (c - lower right) Map of the magnetic field direction
(degrees). (d - lower left) Map of the rotation measure (rad
m$^{-2}$) between 4810 MHz, 1502 MHz and 1452 MHz. All contours are
at $5 \sigma$ at 4810 MHz (1.0mJy beam$^{-1}$)x (-1, 1, 2,
4...,1024). Beam size of 4.5'' x 4.0''.} \label{3c351}
\end{figure}

\begin{figure}
\centerline{\psfig{file=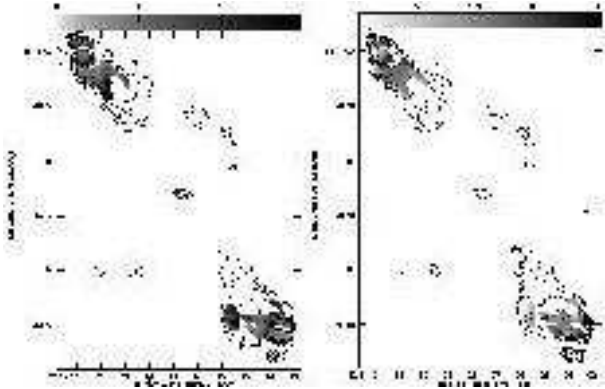,width=8cm}}
\caption{Maps of the radio source {\bf 3C457 }. (a - top left) Map
of the depolarisation between 4710 MHz and 1452 MHz. 
(b - top right) Map of the spectral index between 4710 MHz and 1452 MHz.
 All contours are at $5 \sigma$ at 4710 MHz
(0.3 mJy beam$^{-1}$)x (-1, 1, 2, 4...,1024). Beam size of 5.0'' x 3.0''.} \label{3c457}
\end{figure}

\paragraph*{{\it 3C16:}}(Figure \ref{3c16}) The source 
shows a strong SW lobe, with a
relaxed NE lobe. The SW lobe shows a strong depolarisation
feature that is narrower than the beam size. The strong rotation measure 
feature is evident in the depolarisation map but not the magnetic field map, 
indicating that it is not an error in the fitting program.
No value for the rotation measure was obtained for the NE
lobe because the polarisation observed was too weak.
\paragraph*{{\it 3C42:}}(Figure \ref{3c42})
The core was detected at 4.7 GHz but was absent at the lower
frequencies. The source has been observed to lie in a small
cluster by \citet{vbq00}. \citet{fbp97} observed that the N
hotspot was double but this is not evident in our observations which
can be attributed to the differences in the resolutions of the two observations.
\paragraph*{{\it 3C46:}}(Figure \ref{3c46}) The source has a prominent core at 4710 MHz but 
it is indistinguishable from the extended lobe at 1452 MHz.
\paragraph*{{\it 3C341:}}(Figure \ref{3c341}) The source is a classic double
with a resolved jet-like structure running into the SW lobe. The
jet is more prominent in the higher frequency observations than at
the lower frequencies. 
\paragraph*{{\it 3C351:}}(Figure \ref{3c351}) The source is an extended and
distorted quasar \citep{bhl94}. Both lobes
expand out to envelope the core. The NE lobe is highly extended,
off-axis and shows two very distinct hotspots. The depolarisation
increases towards the more compact SW lobe enforcing the idea that
the environment around the SW lobe is denser, stopping the
expansion seen in the NE lobe. There is evidence of a rotation measure
ridge in the NE hotspots which corresponds to a narrow ridge 
of depolarisation but there is no corresponding shift in the magnetic field map.
\clearpage
\paragraph*{{\it 3C457:}}(Figure \ref{3c457}) The SW lobe shows a prominent double
hotspot. The small compact object just south
of the SW hotspots is most likely an unrelated background object.
The inverted core was observed to be present at all frequencies. This source has
no rotation measure map or magnetic field measure map as we were unable to remove all
n$\pi$ ambiguities from this source. This was due to the small separation of observing 
frequencies around 1.4 GHz and 5 GHz, see Section \ref{blurb}


\begin{figure}
\centerline{\psfig{file=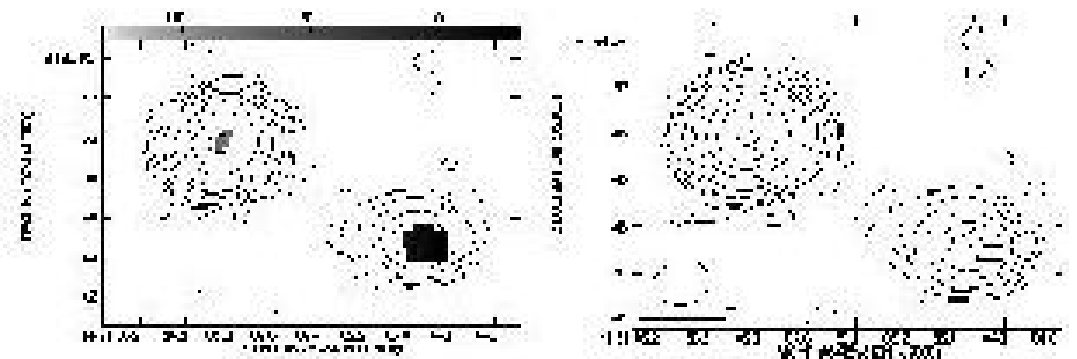,width=8cm}}
\caption{Maps of the radio source {\bf 3C299 }.  (a -  left) Map
of the rotation measure (rad m$^{-2}$) between 4770 MHz, 1502 MHz and
1452 MHz. (b - right) Map of the magnetic field direction
(degrees). All contours are at $5 \sigma$ at 4770 MHz (0.7 mJy
beam$^{-1}$)x (-1, 1, 2, 4...,1024). Beam size of 2.5'' x 2.0''.} \label{3c299}
\end{figure}

\begin{figure}
\centerline{\psfig{file=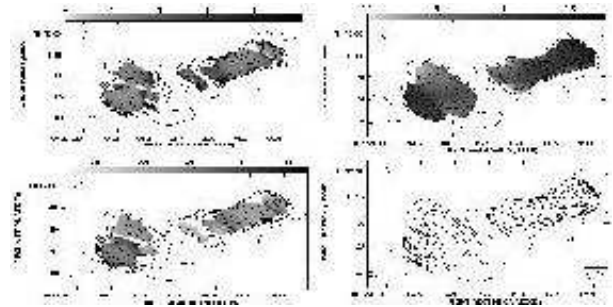,width=8cm}}
\caption{Maps of the radio source {\bf 4C14.27 }. (a - top left)
Map of the depolarisation between 4710 MHz and 1452 MHz. (b - top
right) Map of the spectral index between 4710 MHz and 1452 MHz. (c
- bottom left) Map of the rotation measure (rad m$^{-2}$) between
4885 MHz, 4535 MHz, 1502 MHz and 1452 MHz. (d - bottom right) Map of the
magnetic field direction (degrees). All contours are at $5 \sigma$
at 4710 MHz (0.4 mJy beam$^{-1}$)x (-1, 1, 2, 4...,1024). Beam size of
2.5'' x 2.0''.}
\label{4c14}
\end{figure}

\paragraph*{{\it 3C299:}}(Figure \ref{3c299}) The source is the least luminous in all of the samples.
The source shows a large change in rotation measure between the lobes but the difference is
probably due to the small number of pixels with rotation measure
information in the NE lobe.
\paragraph*{{\it 4C14.27:}}(Figure \ref{4c14})
There is no core detected at any frequency even in a
better quality map by \citet{lp91a}.
\section{Discussion}
\begin{table*}
\centering
\caption{Properties of the sample A radio source components. Errors
are 5\% or less unless stated otherwise. The spectral indices are the
mean values for each component, calculated between approximately 4800
MHz and 1465 MHz. The depolarisation measures are the mean values of
the ratio of the fractional polarisation between approximately 4800
MHz and 1465 MHz for each component. The rotation measures are the
mean values between approximately 4710 MHz, 1665 MHz and 1465 MHz and are quoted in the observer's frame of reference.  The Faraday dispersion, $\Delta$, is given in Equation 5. $\sigma_{RM}$ is the rms in the rotation measure.
All mean values take into account pixels above 5$\sigma_{rms}$ at 4.7 GHz and above 3$\sigma_{rms}$ at 1.4 GHz.}\label{A}
\begin{tabularx}{20cm}{ccccccccccc}
      Source    & Component & Total  & Percentage & Total  & Percentage & Rotation & Spectral & Depolarisation& Average&$\sigma_{RM}$ \\
    & & Flux & Polarisation & Flux & Polarisation & Measure & Index & Measure & $\Delta$\\
    & & 4710 MHz & 4710 MHz & 1465 MHz & 1465 MHz & RM & $\alpha$ & $DM^{4.8}_{1.4}$ \\
         & &(mJy) &\% &(mJy) &\% &(rad m$^{-2}$)&& &(rad m$^{-2}$) &(rad m$^{-2}$)\\
      $6C0943+37$  &  W & 31.0 & 11.8 & 71.7 & 5.4 & 1.6 $\pm 2$ & -0.72 & 2.19 &15.01&16.1\\
   & E  &  42.0 & 2.7 & 136.4 & 0.7 & -19.1 $\pm 6$   & -1.01 & 3.86&19.7  &101.4\\
     $6C1011+36$  &  N  & 44.3 & 7.8 & 119.6 & 5.1 & 30.8 $\pm 5$& -0.88 &1.53& 11.1&12.8
    \\ & S  &  18.7 & 12.2  & 50.9 & 11.5 &12.6 $\pm 4$ & -0.89& 1.06&  4.1&10.1\\
    & Core &3.22 &- &0.7 $\pm 0.5$ &- &- & 1.35& -& - &-\\
      $6C1018+37$  &  NE & 46.4 & 10.0 & 125.8 & 8.1 &0.85$ \pm 3$ & -0.85 & 1.23&7.71& 6.1
    \\ & SW & 28.7  & 7.0 & 76.3 & 2.9 &11.2 $\pm 2$ & -0.84 & 2.41 & 15.9 &2.9 \\
    & Core &0.63 $\pm 0.2$ &-&-&-&-&- & -& -&-\\
      $6C1129+37$  &  NW & 46.5 & 7.3 & 129.1 & 2.9 &-19.3 $\pm 4$& -0.85 &2.51& 16.3 &54.0
    \\  & SE & 73.2 & 16.3 & 215.0  & 3.2 &0.08 $\pm 3$ & -0.90 & 5.09& 21.61&22.0\\
      $6C1256+36$  &  NE & 57.8 & 10.4 & 148.9 & 7.7 & 5.9 $\pm 3$ & -0.79 & 1.35&  9.3&21.8
     \\   & SW &  101.4 & 8.9 & 288.5 & 8.8&15.4 $\pm 3$ & -0.87 & 1.01 & 1.7&14.0\\
      $6C1257+36$  &  NW & 43.5& 17.0 & 102.2  & 13.4 &-115.3 $\pm 9$& -0.71 & 1.27 &  8.3&10.0
    \\  & SE &  20.5 & 9.8& 73.0  & 5.2 & -115.6 $\pm 10$ & -1.06 & 1.88& 13.5&16.0\\
    & Core & 0.29 $\pm 0.08$&-&-&-&-&-&-& -&-\\
      $7C1745+642$ &  N & 23.5 & 9.6 & 64.5 & 5.2&- &-0.86 &1.85& 13.3&-
     \\& Core & 84.1 & 4.9 & 69.4  &2.9  &65.4 $\pm 4$ & 0.16 $\pm 0.1$ & 1.69& 12.3&20.9
    \\ & S & 33.8 & 8.6  & 98.5 & 9.5 &12.8 $\pm 3$ & -0.91 & 0.91& -&3.1\\
      $7C1801+690$ &  N & 8.8 $\pm 3$& 3.0 & 28.2 & 2.7&44.8 $\pm 3$ & -0.97 &1.11&  5.5&16.0
    \\& Core & 79.1 & 1.8 & 78.4 & 4.0 & 30.3 $\pm 3$& 0.007  & 0.45& -&1.80\\
     & S & 28.7 & 10.0& 75.8 & 7.2 & 20.8$\pm 2$ & -0.81 & 1.39&  9.7&10.0\\
      $7C1813+684$ &  NE & 15.0 & 7.7 & 44.8 & 9.5& 13.9$\pm 3$& -0.92 & 0.81&-&85.0
    \\ & SW &  30.2 & 8.4 & 78.1 &7.4 &-68.4$ \pm 7$& -0.79 & 1.14   & 6.1  &41.7  \\
    & Core &3.32 &-&2.65 & -& -& 0.19 $\pm 0.1$ & - & -&-\\
\end{tabularx}
\end{table*}
Table \ref{average} shows the average of the various observed
properties of each of the samples. The average was calculated by
taking the properties of both lobes in each source, averaging them
together and then averaging these values over the sample. Differential
properties, e.g. the difference of rotation measure, dRM, were
calculated by taking the difference of the respective property between
the two lobes of each source and then averaging these values over the
entire sample. These sample-averaged properties give very simple
indicators of trends between samples and hence reveal the strongest
correlations of source properties with redshift and/or radio power.
We will present a more extensive statistical
analysis of these correlations in a forthcoming paper.


\subsection{The location of the Faraday screen}
\label{location}
The observed rotation measure, RM, and degree of polarisation in a
source may be caused by plasma either inside the radio source itself
(internal depolarisation) or by a Faraday screen in between the source
and the observer (external depolarisation). In the latter case the
screen may be local to the radio source or within our own Galaxy, or
both. Only in the case of an external Faraday screen local to the
radio source do our measurements contain information on the source
environment.

The average RM we observe in our sources is consistent with a Galactic
origin \citep{l87}. This is also consistent with the absence of any
significant differences of RM between our samples (see Table
\ref{average}). However, we observe large variations of RM within
individual lobes on small angular scales. These are probably caused by
a Faraday screen local to the source \citep{l87} and therefore must be
corrected for the source redshift by multiplying dRM and $\sigma _{RM}$
by a factor $(1+z)^2$ to allow a valid comparison between sources. The
variation of RM on large angular scales (10s of arcseconds),
i.e. between the two lobes of a source, dRM, may still be somewhat
influenced by the Galactic Faraday screen. Nevertheless, the large
variations of RM found on arcsecond scales measured by $\sigma _{RM}$
suggest an origin local to the source. 
\begin{figure}
\centerline{\psfig{file=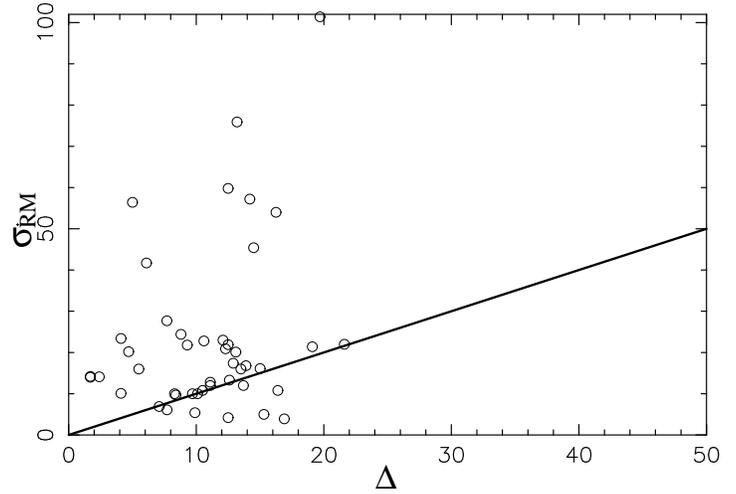,width=9.5cm,angle=270}}
\caption{Plot of the Faraday dispersion, $\Delta$ against the
rms of the rotation measures for each source.}\label{external}
\end{figure}

The observed degree of depolarisation in a source depends on the
distribution of the Faraday depths covered by the projected area of
the telescope beam, i.e. the variation of RM on the smallest angular
scales. We therefore assume that the depolarisation in our sources is
caused by plasma local to the sources. If the depolarisation is caused
by a local but external Faraday screen and the distribution of Faraday
depths in this screen is Gaussian with standard deviation $\Delta$,
then \citep[e.g.][]{burn}

\begin{equation}\label{pp}
m_{\lambda} = m_0 \exp \left\{-2 \Delta ^2 \left[\lambda / \left( 1+z
\right) \right] ^4 \right\},
\end{equation}

\noindent where $m_{\lambda}$ is the percentage polarisation at
observed wavelength $\lambda$ and $m_0$ is the initial percentage
polarisation before any depolarisation. Since we measure $m_{\lambda}$
at two observing frequencies (1.4\,GHz and 4.8\,GHz), we can solve
equation (\ref{pp}) for $\Delta$ as a function of the depolarisation
measure,

\begin{equation}\label{delta}
\centering \Delta=\sqrt\frac{(1+z)^4ln
DM^{4.8}_{1.4}}{2(\lambda_{1.4}^4-\lambda_{4.8}^4)} {\rm
\,rad\,m}^{-2}.
\end{equation}

We can now, for each source, compare the value of $\Delta$ as derived
from the measured depolarisation with the observed rms of the rotation
measure. If the Faraday dispersion is less than the rms of the
rotation measure, then our observations are consistent with an
external Faraday screen \citep{gc91a}.

Figure \ref{external} displays the Faraday dispersion, $\Delta$, for
each lobe of the sources against the rms of the rotation measures
observed. It is evident from the plot that the value of $\sigma_{RM} >
\Delta$ for most components. There are a few sources where this is not
the case.  However, these components belong to sources where the
depolarisation or rotation measure is only determined reliably for a few
pixels so
an accurate value is not obtainable for $\sigma_{RM}$ or $ \Delta$.
There is little correlation between $\Delta$ and $\sigma_{RM}$.  This is
a strong indicator that the Faraday medium responsible for variations
of RM on small angular scales, and thus for the polarisation properties
of our sources, is consistent with being external but local to the
sources.

\begin{table*}
 \centering
  \caption{As Table \ref{A} but for sample B.}\label{B}
\begin{tabular}{ccccccccccc}
      Source    & Component & Total  & Percentage & Total  & Percentage & Rotation & Spectral & Depolarisation & Average& $\sigma_{RM}$ \\
    & & Flux & Polarisation & Flux & Polarisation & Measure & Index & Measure& $\Delta$ \\
    & & 4790 MHz & 4790 MHz & 1465 MHz & 1465 MHz & RM & $\alpha$ & $DM_{1.4}^{4.8}$ \\
         & &(mJy) & \%&(mJy) &\% & (rad m$^{-2}$)&& & (rad m$^{-2}$) & (rad m$^{-2}$) \\
      $3C65$  &  W &  524.0      &19.3 &1683.1 & 5.4 &-82.6 $\pm 7$& -1.0 &3.57& 19.1&21.4   \\
       & E   & 240.9 & 9.2 & 800.4  & 7.2 &-86.1 $\pm 6$& -1.03 & 1.28&  8.4 & 9.7\\
      $3C252$  &  NW  & 178.7 & 6.4 & 592.2 & 5.9 &15.7 $\pm 6$& -1.0 & 1.08 &  4.7&20.2\\
        & SE  &  80.0 & 14.1 &  300.5 & 6.8 &58.5 $\pm 6$& -1.1 & 2.07 & 14.5 &45.4\\
    & Core &1.98 &-&1.33 &-& -& 0.33&-& -&-\\
      $3C267$  &  E & 184.0 & 8.9 & 745.7 & 6.8 &-9.6 $\pm 3$& -1.17 &1.31&  8.8&24.4 \\
    & Core & 1.87 $\pm 1$& - & 1.05 $\pm 2$& - & - & 0.48 & -& - &-\\
    & W & 479.2  & 3.5 & 1294.6 & 3.3 &-21.5$\pm 3$& -0.83 & 1.06&  4.1&23.4    \\
      $3C280$  & E& 326.0 & 8.0 & 1219.2 & 4.4 &-37.7$\pm 8$& -1.01& 1.82& 13.1&20.1\\
      & W & 1289.2 & 10.0 & 3191.6  &6.5&-7.5 $\pm 4$& -0.76 & 1.54 & 11.1&1.5 \\
      $3C324$  &  NE & 432.6 & 9.3 & 1525.2 & 5.6& 22.1 $\pm 4$& -1.05& 1.66&12.1&23.0 \\
    & SW &  166.0 & 7.8 & 651.8 & 4.0 &43.0 $\pm 5$& -1.14 & 1.95& 13.9&16.8 \\
      $4C16.49$ &  N  &107.2& 8.3  & 307.2 & 13.0 &-4.3 $\pm 5$& -0.88 & 0.64&- &56.4\\
    & Jet&  8.5 $\pm 4$&14.5& 45.9  & 7.2 &30.1 $\pm 4$& -1.41& 2.01& 14.2 &57.2\\
                &Core & 9.64 $\pm 2$& 2.7  & 58.2 &3.5 & - & -1.50 & 0.77& - &-\\
     & S  & 142.3 &7.0&695.3  &6.4 & 0.9 $\pm 3$& -1.32 & 1.09 & 5.0 &56.4\\
      $3C68.1    $ & N  &667.2 & 8.6& 1767.4&7.0 &-26.6$\pm 5$& -0.82 &1.23&  7.7&27.7 \\
    &  S &36.8 &11.4   &134.6 &  6.6 & 57.9 $\pm 2$ & -1.08 & 1.73 & 12.5 &59.8\\
      $3C265     $ & NW &224.0&10.0& 478.7&5.8&42.2 $\pm 6$& -0.62 & 1.72&12.5&21.9\\
    &    SE& 318.9&6.2  &835.0 &4.2 & 32.8$\pm 3$& -0.78 & 1.48 &10.6 &22.8  \\
      $3C268.1   $ &  E   & 262.3& 5.0 &816.4 & 3.4 & 21.7 $\pm 5$&-0.92&1.47& 10.5&10.8\\
    &  W   &2296.6 &4.7&4699.0&  2.7  & 26.8 $\pm 6$ & -0.58 & 1.74      & 12.6 &13.3 \\

\end{tabular}
\end{table*}
\subsection{Trends with redshift and radio power}

\subsubsection{Spectral index}

We found no significant correlation between spectral index and
redshift or radio power as suggested by others
\citep{on89,vv72,athreya}, which used much larger samples. This
suggests the trend may be present but because of the fact that our
sample is small we do not find any significant trend.  However there
are trends found with the difference in spectral index between the two
lobes.  Sample B shows a larger average difference in the spectral
indices between the two lobes of a given source than the other samples
(Sample A: 2.5$\sigma$, sample C: 3.3$\sigma$). Sample B contains the
sources with the highest radio power and so we find that in our
samples the difference in spectral index increases with radio power
rather than with redshift.  The trend of the difference in the
spectral index between the two lobes may be related to the extra
luminosity of the 3CRR hotspots compared to the 6C/7C hotspots (see
Section \ref{obs}). On average we find that the hotspots of our
sources have shallower spectral indices. The average spectral index
integrated over the entire sources will therefore depend on the
fraction of emission from the hotspots compared to the extended lobe,
thus creating the observed trend.
\begin{table}
\centering \caption{Reduced $\chi^2$ values for the rotation measure fits for
Figures \ref{4lambda} to \ref{6lambda}.}\label{values}
\begin{tabular}{ccccccc}
Source &Lobe&n=-1& n=0 & n=1 &Symbol\\
6C1256+36&S&382.&51.&1.91&x\\
&&981.6&1.23&976.5&o\\
4C16.49&N&1.6&669.&2917.&o\\
&&2772.&90.1&1.7&x\\
4C16.49&S&151.9&0.9 &62.&o\\
&&45.4&0.4  &34.8  &x\\
\end{tabular}
\end{table}

\subsubsection{Rotation measure}
As mentioned above, we find no significant difference in the average
RM between our samples. Similarly, there is no statistically
significant trend of the difference of RM between the two lobes of
each source with redshift or radio power. This is consistent with a
Galactic origin of the RM properties of the sources on large
scales. On small angular scales the variation of RM as measured by its
rms variation $\sigma _{RM}$ shows a significant trend between the low
redshift sample (C) and the high redshift samples (A: 2.6$\sigma$, B:
3.3$\sigma$). The exclusion of 3C457 due to $n\pi$-ambiguities in the
rotation measure (see section \ref{blurb}) from the rotation measure
averages could bias the sample C results towards a small value of
$\sigma _{RM}$. In principle, similar ambiguities could also affect
other sources in sample C. However, as stated above we find the
rotation measure to vary smoothly across the other sources in this
sample. The exclusion of one source will not remove the trend we find
here.  There is no significant difference between the low power
sample A compared with the high power sample B. This suggests that at
least the range of rotation measures on small angular scales in a
source does depend on the source redshift and not the source radio
power.  In a recent study by \citet{pent} rotation measure was also
found to be independent of source luminosity and size but dependent on
the redshift of the source.

\begin{figure*}
\centerline{\psfig{file=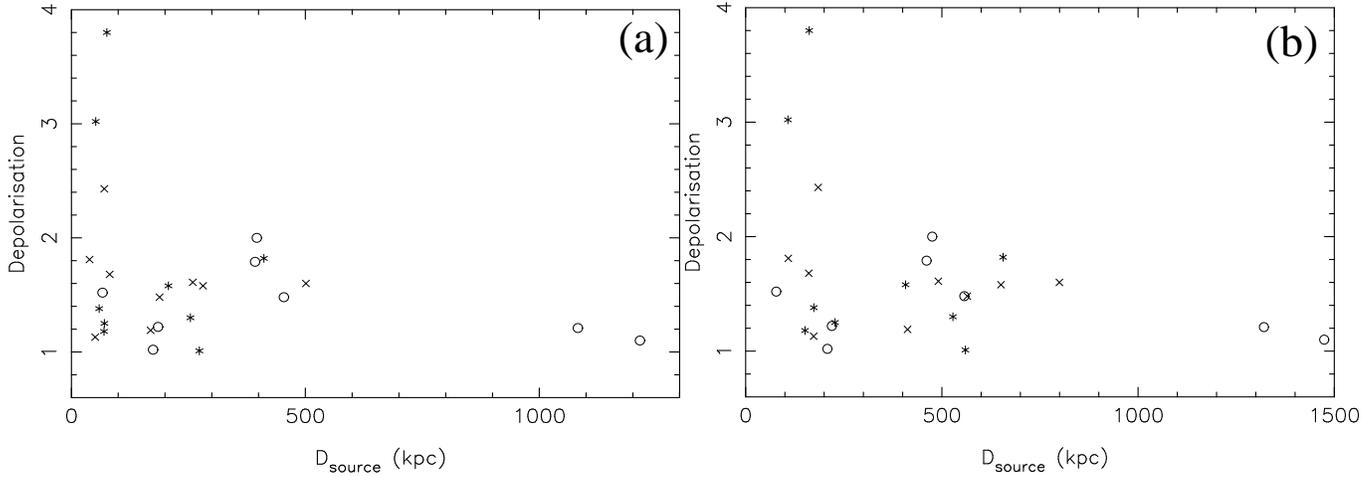,width=18cm}}
\caption{Average depolarisation against projected linear size
(kpc) for all 3 samples. Symbols as in Figure \ref{pzplot}. Figure (a) assumes $H_o= 50$
kms$^{-1}$Mpc$^{-1}$, and $\Omega_{m} = 0.5$, $\Omega_{\Lambda} = 0$. Figure (b) assumes $H_o= 50$
kms$^{-1}$Mpc$^{-1}$, and $\Omega_{m} = 0.35$, $\Omega_{\Lambda} = 0.65$.}\label{depolsize}
\end{figure*}

\subsubsection{Percentage polarisation}

The flux observed from both lobes of all sources was found to be
polarised at levels greater than 1\%, the only exceptions being
6C0943+39 and 3C299. At lower redshifts the polarisation exhibits the
largest range from 0.8\% in 3C299 to 28.0\% in 3C341. This range is
not evident in either of the other two samples. Statistically the
sources at low redshift (sample C) are slightly more polarised than
those at high redshift both at 1.4\,GHz (sample A: 2.1$\sigma$, sample
B: 2.3$\sigma$) but at 4.8\,GHz there is no significant difference. 
There is no trend observed comparing the low radio power
sources (sample A) with the high power objects (sample B). This
suggests that percentage polarisation decreases for increasing
redshift but is less dependent on radio power.

So far we considered percentage polarisations measured in the
observing frame. We argue in section \ref{location} that variations of
the rotation measure on small angular scales which determine the
degree of polarisation are caused in Faraday screens local to the
sources. The trend with redshift may therefore simply reflect the
different shifts of the observing frequency in the source restframe
for sources at low and high redshift. Using Burn's law in the form of
equations (\ref{pp}) and (\ref{delta}) we can determine for each
source the percentage polarisation expected to be observable at a
frequency corresponding to a wavelength of 5\,cm in its restframe. The
results for individual sources are presented in Table \ref{hmm} and
the sample averages are summarised in Table \ref{averagez=1}. Again
sources at low redshift (sample C) are slightly more polarised than
sources at high redshift but only 
sample B (2.1$\sigma$) shows a result that is marginally significant.  As before,
there is no trend with radio power, indicating that the trend with
redshift, although weak is dominant. This is not caused by pure Doppler shifts of the
observing frequencies.

\begin{table*}
 \centering
  \caption{As Table \ref{A} but for sample C. Rotation measure is between 4800, 4500, 1502 and 1450}\label{C}
\begin{tabular}{ccccccccccc}
      Source    & Component & Total  & Percentage & Total  & Percentage & Rotation & Spectral & Depolarisation& Average&$\sigma_{RM}$ \\
    & & Flux & Polarisation & Flux & Polarisation & Measure & Index & Measure&$\Delta$ \\
    & & 4710 MHz & 4710 MHz & 1452 MHz & 1452 MHz & RM &$\alpha$  & $DM_{1.4}^{4.8}$ \\
         & &(mJy) &\% &(mJy) &\% &(rad m$^{-2}$)&&& (rad m$^{-2}$)& (rad m$^{-2}$) \\
      $3C42$  & NW & 353.9 & 12.5 & 999.7 & 12.3 &  -2.4$\pm 5$& -0.87 & 1.02 & 2.4&14.1\\
       & SE & 450.6   & 7.5 &1266.3 & 7.4 &  5.0$\pm 5$& -0.86 & 1.01 & 1.7&14.2\\
      $4C14.27$  &  NW  & 107.0 & 12.4 & 368.2 & 8.8 &-13.0$\pm 4$& -1.06 & 1.41& 9.9&5.4
    \\& SE  &  124.8& 9.6  &  489.3 & 9.3 &  -17.3 $\pm 5$& -1.17& 1.03& 2.9&4.8   \\
      $3C46 $  & NE & 162.7 &16.2 & 488.0 &12.8 &-4.8 $\pm 5$& -0.94 & 1.27&16.9&3.9
    \\&SW & 173.7  &13.7 & 506.4  & 12.0 &-2.9 $\pm 1$& -0.91 & 1.14 & 12.5&4.2   \\
    & Core &2.54 &-&7.94 &-& -& -0.97&-& -&-\\
      $3C457$  &NE& 208.6 &15.8 & 692.0  &15.7 &-&-1.02 & 1.01 & -  &-   \\
       &SW & 290.5  & 11.9 & 898.9 & 10.0&-          & -0.94 &1.19& -   &-    \\
    &Core & 3.03&-&2.63 & - &-&0.12&-& -&-\\
      $3C351$  &  NE & 971.7 & 6.8 & 2327.9 & 8.1 & 1.00 $\pm 5$ & -0.73 & 0.84 & -&13.0\\
    &Diffuse & 122.2 & 26.4 & 421.3 & 13.7 & -8.7 $\pm 1$ & -1.03 &  1.93 & 13.7&12.0\\
    &Core & 18.3 & 18.6 & 46.2 & 8.2 & -0.53$\pm 2$ &-0.77&2.27 & 15.3&5.0\\
    & SW &  77.1 & 22.8 & 235.7 & 8.9 &4.4 $\pm 3$&-0.93&2.56& 16.4&10.8\\
      $3C341  $ &  NE &123.8&28.0  & 336.4 & 19.6 &20.3 $\pm 5$& -0.85 & 1.43& 10.1&10.0
        \\ & SW  & 265.9 &26.4&862.6  &17.2 &18.2 $\pm 5$& -1.0  &1.53 & 11.1&12.0\\
      $3C299     $ & NE &876.5& 0.79$\pm 0.5$& 2592.4&0.43 $\pm 0.3$&-126.3 $\pm 8$& -0.93 & 1.84& 13.2&75.9
    \\&  SW&53.7 &3.1   &122.6&  2.6 &16.0 $\pm 5$& -0.71 & 1.19& 7.10& 6.9\\
      $3C16      $ & NE &22.1 $\pm 3$ &10.8  &60.9  & 4.9 & -&-0.87& 2.2 & 15.0&- \\
    & SW   & 484.9& 14.7& 1510.9 &8.2 &-4.3 $\pm 3$& -0.97 &1.79 &  12.9& 17.4 \\
\end{tabular}
\end{table*}

\subsubsection{Depolarisation}

Comparing the average depolarisation of individual samples with each
other we find only very weak trends with redshift or radio power. When
the samples are averaged together we find a trend in depolarisation
with redshift but none with radio power. Samples A+C (low radio power)
are statistically identical to sample B (high radio power), suggesting
that there is no trend with radio power in our sources. By
considering the averaged depolarisation of samples A+B (high redshift)
compared with that of sample C (low redshift) there is a weak trend
with redshift which is echoed in the dDM values. However both
results are not significant. This may confirm the results of
\citet{kcg72}: redshift is the dominant factor compared to radio power
in determining the depolarisation properties of a source.
\begin{table*}
\centering \caption{Mean properties of the sources averaged over each
sample with the associated error. Differential properties
(dDM,d$\alpha$,dRM) are derived by taking the difference of the
respective property between the two lobes of each source and then
averaging this difference over each sample, dRM is in the source frame.$\sigma_{RM}$ is the rms of 
the RM over the source, in the source reference frame. Properties in italics are in the source frame of 
reference.}\label{average}
\begin{tabular}{cccccc} 
Property & sample& sample& sample& sample& sample \\
&	A& B&C&A+B&A+C\\
Average z&1.06$\pm 0.04$ & 1.11$ \pm 0.05$ & 0.41 $\pm 0.01$ & 1.08 $\pm 0.03$ & 0.75 $\pm 0.08$ \\
Average P$_{151}$&$(1.02 \pm 0.06)\times 10^{27}$&$(1.01 \pm 0.07)\times 10^{28}$&$(8.02 \pm 0.40)\times 10^{26}$&
	$(5.57 \pm 1.16)\times 10^{27}$&$(8.97 \pm 0.47)\times 10^{26}$\\
Average D&220.34$\pm 46.40$ & 262.69$ \pm 55.41$ & 399.57 $\pm 122.50$ & 241.51 $\pm 35.43$ & 304.69 $\pm 64.50$ \\
Average DM&1.82 $\pm 0.32$ & 1.61$ \pm 0.13$ & 1.42 $\pm 0.12$ & 1.71 $\pm 0.17$ & 1.63 $\pm 0.18$ \\
 dDM& 0.93 $\pm 0.26$ & 0.61 $\pm 0.22$ & 0.43 $\pm 0.18$& 0.77$ \pm 0.17$ & 0.69 $\pm 0.17$ \\
Average $\alpha$ & -0.87 $\pm 0.01$ & -0.92 $\pm 0.06$ & -0.94 $\pm 0.03$ &-0.90$ \pm 0.03$ &-0.90 $\pm 0.02$ \\
 d$\alpha$& 0.13 $\pm 0.04$ &0.23 $\pm 0.04$ & 0.10 $\pm 0.03$& 0.18$ \pm 0.03$ & 0.12 $\pm 0.02$  \\
Average PF$_{1.4}$&6.29$\pm 0.86$ & 5.88$ \pm 0.64$ & 9.89 $\pm 1.75$ & 6.08 $\pm 0.52$ & 7.98 $\pm 1.01$ \\
Average PF$_{4.8}$&8.91 $\pm 1.03$ & 8.76$ \pm 0.89$ &13.31 $\pm 2.48$ & 8.83 $\pm 0.66$ &10.98 $\pm 1.36$ \\
Average RM&  29.03 $\pm12.81$ & 26.41 $\pm 7.74$ &16.24 $\pm 8.33$ & 27.64$ \pm  7.05$ &23.55 $\pm 8.07$ \\
{\it dRM}&97.85 $\pm36.41$ &115.96 $\pm42.81$ &50.77$\pm43.29$&107.44$\pm27.62$ &77.68 $\pm27.54$  \\
{\it $\sigma_{RM}$}&115.62 $\pm33.43$ & 122.94 $\pm28.48$ & 29.45 $\pm 9.98$&119.49 $\pm21.11$&78.69 $\pm22.36$\\
\end{tabular}
\end{table*}

Analogous to the discussion above on percentage polarisation, the
cosmological Doppler shifts of the observing frequencies influence the
trend of DM with redshift, and in fact the true trend is stronger
than that naively observed. To demonstrate this we use equation
(\ref{delta}) to derive the standard deviation of Faraday depths,
$\Delta$, for each source. By setting $z=1$ we then rescale all the
depolarisations DM$^{4.8}_{1.4}$ to the same redshift. This allows
all 3 samples to be compared without any bias due to pure redshift
effects (see Tables
\ref{hmm} and \ref{averagez=1}).  If there was no intrinsic difference
between the high-redshift and low-redshift samples then we would
expect these values to be consistent with each other. This is
evidently not the case. The high redshift samples are, on average
significantly more depolarised (sample A: 2.2$\sigma$, sample B:
3.2$\sigma$) than their low-redshift counterparts (sample
C). Comparing sample A with sample B we find no trend with radio
power. However, a note of caution must be issued as the corrections
applied use Burn's law and may actually be too large. Considering the
precorrected and the corrected values together it is obvious that
there is a connection between redshift and depolarisation but there is
no significant trend of depolarisation with radio power. There is also
a connection between the difference in the depolarisation, dDM, and
the redshift of the source, but no significant trend of the difference
in the depolarisation with the radio power of the source. As noted in
Section \ref{blurb} regions with signal-to-noise $< 3\sigma$ were
blanked in the map production. In individual sources blanking of low
S/N regions in the polarisation maps will cause the measured
depolarisation to be underestimated. Sources in sample A are more
affected by this problem than objects in the other samples. Therefore
we probably underestimate the average depolarisation in sample A
implying that the trend with redshift could be even stronger than our
findings suggest.

The trends of percentage polarisation and of depolarisation with
redshift are probably related in the sense that a lower degree of
depolarisation at low redshift also leads to a higher observed degree
of polarisation. Clearly a variation of the initial polarisation,
$m_0$, with redshift would lead to variations of the observed
$m_{\lambda}$ independent of the properties of any external Faraday
screen. Therefore both trends could also be caused by a significantly
higher level of $m_0$ of the sources at low redshift (sample C). Using
equation (\ref{pp}) we find $m_0=9.3\pm1.1$ for sample A,
$m_0=9.1\pm1.0$ for sample B and $m_0=13.5\pm2.5$ for sample C. The
uncertainties associated with the use of Burn's law in extrapolating
from our observations to $\lambda =0$ are large. There is no
difference between average initial polarisation of sources in samples
A and B. The difference found for $m_0$ at low redshift (sample C)
compared to high redshift (samples A and B) is small compared to the
difference found for the depolarisation comparing the same
samples. This suggests that the differences in percentage polarisation
and depolarisation are due to variation with redshift of the Faraday
screens local to the sources rather than to differences in the initial
degree of polarisation. However note, that the variation of $m_0$ is
not significantly smaller than the trend of percentage polarisation
with redshift.

\begin{table}
\centering \caption{Recalculated average depolarisation and dDM for each
source if it was located at z=1 and average percentage polarisation of all sources if it
was emitted at 5cm in the rest frame.}\label{hmm}
\begin{tabular}{cccccc}
Source & z &	$\Delta_z$	&DM$_{z=1}$& dDM$_{z=1}$& Polarisation\\
& & & && $\lambda_{rest}=5cm$\\ 
6C0943+37 &1.04 & 74.12 &	 3.31 & 1.98 &3.1 \\
6C1018+37 &0.81 & 42.95 &	 1.49 & 0.52&8.32\\
6C1011+36 & 1.04 & 36.11 & 1.33 & 0.66&5.63\\
6C1129+37 &1.06 & 83.07 &	 4.49& 3.43&3.09\\
6C1256+36 &1.07 & 29.53 & 	1.21 & 0.40&8.26\\
6C1257+36 &1.00 & 45.51 & 	1.58 & 0.61&9.39\\
7C1801+690 &1.27 & 41.24 &	 1.45 & 0.54&3.33\\
7C1813+684 &1.03 &  6.96 &	 1.01& 0.25&8.47\\ \vspace{0.3cm}
7C1745+642 &1.23 & 47.81 &	 1.65& 1.36&7.32\\ 
 3C65 &1.18 & 75.33 & 		3.50 & 4.61&6.26\\
3C252 &1.11 & 50.66 & 		1.76 & 1.36&6.36\\
3C267 &1.14 & 32.34 & 		1.26 & 0.35&5.04\\
3C280 &1.00 & 48.79 & 		1.68 & 0.28&5.48\\
3C324 &1.21 & 54.34 & 		2.42 & 0.58&4.74\\
3C265 &0.81 & 35.56 & 		1.37 & 0.14&5.05\\
3C268.1 &0.97 & 42.13 & 	1.57 & 0.25&3.07\\
 3C68.1 &1.24 & 45.11 & 	1.85 & 0.98&10.2\\  \vspace{0.3cm}
 4C16.49 &1.29 & 31.04 & 1.23 & 0.47&6.76 \\
 3C42 &0.40 &  4.59 & 		1.00 & 0.01&9.85\\
 4C14.27 &0.39 & 14.34 &	 1.05 & 0.08&9.08\\
 3C46 &0.44 & 15.06 & 		1.05 & 0.03&12.45\\
3C457 &0.43 & 10.50 &		 1.03 & 0.04&12.88\\
3C351 &0.37 & 23.82 & 		1.14 & 0.15&8.59\\
3C341 &0.45 & 21.91 & 		1.11 & 0.02&18.57\\
 3C16 &0.41 & 27.54 & 		1.19& 0.06&6.64\\
3C299 &0.37 & 20.20 & 1.10 & 0.1&1.52\\
\end{tabular}
\end{table}

Burn's law predicts a steep decrease of percentage polarisation with
increasing observing frequency. Although the decrease may well be
`softened' by geometrical and other effects in more realistic source
models \citep{laing}, the value of DM may be small for sources in
which our observing frequencies are lower than the frequency at which
strong depolarisation occurs. For such sources we expect to
measure a low value of DM associated with a low percentage
polarisation. \citet{ag95} show that one of our sources, 3C299 (sample
C), is strongly depolarised between 1.6\,GHz and 8.4\,GHz with almost
all of the depolarisation taking place between 4.8\,GHz and
8.4\,GHz. We measure only a very low percentage polarisation for this
source and the average value of DM$^{4.8}_{1.4} = 1.5$ is also lower
than DM$^{8.4}_{1.6} \sim 4$ as measured by \citet{ag95}. If many of
our sources at low redshift were affected by strong depolarisation at
frequencies higher than 4.8\,GHz, then this may cause the trend of DM
with redshift noted above. The absence of any other sources with very
low degrees of polarisation combined with a low value for DM, at least
in the low redshift sample C, argues against this bias. In fact,
\citet{tab} show that our sources 3C42, 3C46, 3C68.1, 3C265, 3C267,
3C324 and 3C341 depolarise strongly only at frequencies lower than our
observing frequencies. 3C16, 3C65, 3C252, 3C268.1 and 3C280 do
depolarise strongly between 1.4 GHz and 4.8 GHz. In all the sources
mentioned in the \citet{tab} none have strong depolarisation at frequencies higher than
4.8 GHz.  There is no information for 4C14.27, 4C16.49, 3C351, 3C299 and
3C457. Those sources which do depolarise strongly between our observing
frequencies, 1.4 and 4.8\,GHz, could in principle have inaccurate values
of the rotation measure because of this. However, the pixels containing
most depolarised regions of the source are likely to have been blanked
because of insufficient signal--to--noise in their polarisation at
1.4\,GHz; the rotation measure is determined from the unblanked (less
depolarised) regions of the source, and will therefore be reliable.

\begin{table*}
\centering \caption{Average DM and dDM now shifted so that all the measurements are taken at z=1.
Average polarisation of all the sources, shifted to a common rest frame wavelength of 5cm using
Equation \ref{delta}}\label{averagez=1}
\begin{tabular}{cccccc} 
Property & sample& sample& sample& sample& sample \\
&	A& B&C&A+B&A+C\\
Average DM$_{z=1}$&1.95 $\pm 0.39$ & 1.85$ \pm 0.24$ & 1.08 $\pm 0.02$ & 1.90 $\pm 0.22$ & 1.54 $\pm 0.23$ \\
 dDM$_{z=1}$& 1.08 $\pm 0.34$ & 1.00 $\pm 0.47$ & 0.06 $\pm 0.02$& 1.04$ \pm 0.28$ & 0.60 $\pm 0.22$ \\
PF$_{\lambda_{rest}=5cm}$&6.32 $\pm 0.86$ & 5.88$ \pm 0.65$ & 9.82 $\pm 1.85$ & 6.10 $\pm 0.53$ & 7.97 $\pm 1.04$ \\
\end{tabular}
\end{table*}

 \citet{strom,sj88,prm89,ics98} find an anti-correlation of
depolarisation with linear size. Thus our findings could be the result
of our sample C containing larger sources than samples A and B.
Figure \ref{depolsize} shows that there is a weak anti-correlation
between physical source size and depolarisation. A Spearman rank test
gives a confidence level of around 90\% for this
anti-correlation. However, this trend is due to the two largest
sources in sample C, 3C46 and 3C457. Removing these sources yields an
average depolarisation measure of $1.50\pm0.15$ for sample C, which is
not significantly different from the average with these two sources
included. Thus we can rule out the possibility that the larger
depolarisation at high redshift is caused by selecting preferentially
small sources at low redshift.

\citet{lp91} observed depolarisation to correlate with spectral index
but there are no obvious correlations in any of our samples when we
examined the sources individually or as a combined sample. 

8 sources show dDM $\ge 1$ over the source and a further 6 show $0.5
\le {\rm dDM} <1$. This implies that 14 sources out of 24 sources, for
which a depolarisation measurement exists in both lobes, show a
significant asymmetry in the depolarisation of their lobes.  We would
expect a proportion of the sources to be observed at angles
considerably smaller than 90$^{\circ}$ to our line-of-sight. Therefore
it is not surprising that so many sources are found to be candidates
for the Laing-Garrington effect. 

Several sources show signs of depolarisation shadows in at least one
of the lobes. Of these, 6C1256+36 was observed to lie in a cluster by
\citet{reh98}, as were 3C65 and 3C324 \citep{b00}.  3C324 has been
observed by \citet{brb98} in the sub-mm ($850\mu$m) and found a large
dust mass centred around 3C324. The host galaxy was shown to be the
cause of the very strong depolarisation, \citep{bcg98}.

\subsubsection{Summary}

From the trends listed above, we can conclude that the variations of
RM on small angular scales and the associated depolarisation of the
radio emission are likely caused by a Faraday screen which is external
but local to the source. This implies that any trends with radio power
and/or redshift reflect changes of the source environment depending on
these quantities. We find that percentage polarisation decreases with
redshift while depolarisation increases with redshift. According to
Burn's law \citep{burn}, this implies an increase in the source
environments of either the plasma density or the magnetic field
strength or both with redshift. Such an interpretation is also
supported by the increased depolarisation asymmetry of sources at
high redshift compared with their low redshift counterparts. Note
however, that the trends reported here are based on averaging the
observed properties of sources within three samples and that some of them
are not highly significant. A more detailed statistical analysis of our
observations in a forthcoming paper will help to test these crude
trends.

\section{Conclusions}
In this paper we present the complete data set of our three samples of
radio galaxies and radio-loud quasars. The three samples were
defined such that two of them overlap in redshift and two have similar
radio powers. Thus we are able to study the effects of redshift and
radio power on various source properties.

Even without a formal analysis of the correlation between source
properties, some general trends are already discernible.  There is
little correlation between $\Delta$ and $\sigma_{RM}$, suggesting that
the Faraday medium responsible for variations of RM on small angular
scales, is consistent with being external but local to the
sources. There is also little correlation between rotation measure and
redshift or radio power which is consistent with a Galactic origin of
the RM properties of the sources on large scales.  However, we find
that the rms fluctuations of the rotation measure correlate with
redshift but not radio power to a confidence level of $> 99.9\%$,
determined in the sources' frame of reference.

We find that the polarisation of a source anti-correlates with its
redshift but is independent of its radio power, resulting in the low
redshift sample having much higher degrees of polarisation, in
general.  We also detect higher degrees of depolarisation in the high
redshift samples (A and B) compared to sources at lower redshift
(sample C). This suggests that depolarisation is correlated with
redshift. These two results are probably related in that lower
depolarisation at low redshift leads to both lower depolarisation
measurements and also higher degrees of observed polarisation. 

Our findings on the rotation measurements and polarisation properties
of our sources are indicative of an increase of the density and/or the
strength of the magnetic field in the source environments with
increasing redshift.

We find a number of possible depolarisation shadows, mostly in sources known
to be located in cluster environments.

We find no correlation between the spectral index and redshift or
radio power. However, we do find the the difference in the spectral
index, across individual sources, increases for increasing radio
power of the source and also with increasing redshift of a
source.

A more detailed investigation into these findings will be presented in a
forthcoming paper.

\section*{Acknowledgements}
We would like to thank Mark Lacy for his radio data of the 7C sources
and helpful advice. We are grateful to VLA archives for providing us
with the archival data. We also thank our referee, J.P. Leahy, for
many helpful comments. J.A. Goodlet would like to thank PPARC for
financial support in the form of a studentship. P.N. Best would like
to thank the Royal Society for generous financial support through its
University Research Fellowship scheme.

\bibliography{jag}
\bibliographystyle{mn2e}

\label{lastpage}

\end{document}